\documentclass[journal,10pt]{IEEEtran}
\usepackage{stfloats}
\usepackage{amsmath}
\usepackage{amssymb,latexsym}
\usepackage{graphicx}
\usepackage{color,cite,times}
\usepackage{epstopdf}
\usepackage{algorithmic}
\usepackage[ruled,linesnumbered] {algorithm2e}
\usepackage{booktabs}
\usepackage{bbm} % the Letter hollowing
\usepackage{url}
\usepackage{fancyhdr}
\usepackage{cleveref}
\usepackage{verbatim}
\usepackage{multirow}
\usepackage{makecell}
\usepackage{graphicx}
\usepackage{indentfirst}%大括号编号
\usepackage{cases}
\newcommand{\RNum}[1]{\uppercase\expandafter{\romannumeral #1\relax}}
\usepackage{colortbl} %表格颜色填充
\usepackage{extarrows}
\usepackage{threeparttable}
\usepackage{diagbox}
\usepackage{pifont}
\usepackage{cancel}
\usepackage{framed}
\usepackage{overpic}

\usepackage{subfigure}

\usepackage[framemethod=tikz]{mdframed}
\usepackage{lipsum}

\newcommand{\bsm}{\boldsymbol}

\newcommand{\mbf}{\mathbf}
\newcommand{\diag}{\mathrm{diag}}

\definecolor{gray}{RGB}{192,192,192}

\begin{document}
\title{Joint Activity Detection and Channel Estimation for Massive Connectivity: Where Message Passing Meets Score-Based Generative Priors}

\author{Chang Cai, \IEEEmembership{Member, IEEE,}
	Wenjun Jiang, \IEEEmembership{Member, IEEE,}
	Xiaojun Yuan, \IEEEmembership{Fellow, IEEE,} \\
	and Ying-Jun Angela Zhang, \IEEEmembership{Fellow, IEEE} 
		\thanks{
			This work was supported in part by the China Postdoctoral Science Foundation under Grant GZC20252291 and Grant 2025M783500;
			in part by the National Natural Science Foundation of China under Grant 62571087;
			in part by the General Research Fund from the Research Grants Council of Hong Kong under Project 14214122, Project 14202723, and Project 14207624;
			in part by the Area of Excellence Scheme Grant from the Research Grants Council of Hong Kong under Project AoE/E-601/22-R;
			and in part by the NSFC/RGC Collaborative Research Scheme from the Research Grants Council of Hong Kong under Project CRS\_HKUST603/22 and Project CRS\_HKU702/24.
			\textit{(Corresponding author: Xiaojun Yuan.)}
			
			Chang Cai was with the Department of Information Engineering, The Chinese University of Hong Kong, Hong Kong, and is now with the Department of Electrical and Electronic Engineering, The University of Hong Kong, Hong Kong (e-mail: changcai@hku.hk).
			
			Wenjun Jiang and Xiaojun Yuan are with the National Key Laboratory of Wireless Communications,
			University of Electronic Science and Technology of China, Chengdu 611731, China (e-mail: wjjiang@std.uestc.edu.cn; xjyuan@uestc.edu.cn). 
			
			Ying-Jun Angela Zhang is with the Department of Information Engineering, The Chinese University of Hong Kong, Hong Kong (e-mail: yjzhang@ie.cuhk.edu.hk).
			
		}
	}
\maketitle
\begin{abstract}
	Massive connectivity supports the sporadic access of a vast number of devices without requiring prior permission from the base station (BS).
	In such scenarios, the BS must perform joint activity detection and channel estimation (JADCE) prior to data reception.
	Message-passing algorithms have emerged as a prominent solution for JADCE under a Bayesian inference framework.
	The existing message-passing algorithms, however, typically rely on some hand-crafted and overly simplistic priors of the wireless channel, leading to significant channel estimation errors and reduced activity detection accuracy.
	In this paper, we focus on the problem of JADCE in a multiple-input multiple-output orthogonal frequency division multiplexing (MIMO-OFDM) grant-free random access network.
	We propose to incorporate a more accurate channel prior learned by score-based generative models (a.k.a. diffusion models) into message passing, so as to push towards the performance limit of JADCE.
	Specifically, we develop a novel turbo message passing (TMP) framework that models the entire channel matrix as a super node, rather than factorizing it element-wise.
	This design enables the seamless integration of score-based generative models as a minimum mean-squared error (MMSE) denoiser.
	The variance of the denoiser, which is essential in message passing, can also be learned through score-based generative models.
	Our approach, termed score-based TMP for JADCE (STMP-JADCE), takes full advantages of the powerful generative prior and, meanwhile, benefits from the fast convergence speed of message passing.
	Numerical simulations show that STMP-JADCE drastically enhances the activity detection and channel estimation performance compared to the state-of-the-art baseline algorithms.
\end{abstract}


\section{Introduction} 
Evolved from massive machine-type communication (mMTC) in fifth-generation (5G), 
massive communication (a.k.a. massive connectivity) has been identified as one of the six key usage scenarios in sixth-generation (6G) wireless networks \cite{ITU2023}.
A unique feature of massive communication is the sporadic transmission of short packets by massive machine-type devices, e.g., Internet-of-Things (IoT) devices, where only a small fraction of devices are active in packet transmission at any given time instant \cite{mMTC2016comm}.
To support massive connectivity of machine-type devices, conventional grant-based random access schemes require each device to obtain a grant from the base station (BS) before transmission.
The uplink grant request involves a complex handshaking process, which may result in significant signaling overhead and access latency \cite{latency2017comm, chenxiaoming2021jsac}, especially in scenarios with massive connected devices.

In contrast, grant-free random access \cite{mMTC2016comm, latency2017comm, chenxiaoming2021jsac, chenzhilin2018tsp, liangliu2018tsp} enables active devices to access the uplink channel without waiting for BSs' permission, making it a more appealing solution for massive connectivity.
Specifically, each active device first sends a unique pilot sequence to the BS, and then transmits payload data directly.
Since the device activity is unknown at the BS beforehand, the BS must first detect the active devices and estimate their channels prior to payload data reception.
However, due to the large number of devices and the limited resources for pilot transmission, it is impractical for the pilot sequences to be mutually orthogonal.
As a result, the received pilots suffer from significant co-channel interference, which necessitates the development of advanced algorithms for accurate activity detection and channel estimation.

Due to the sporadic traffic of massive devices, activity detection and channel estimation can be formulated jointly as a compressive sensing (CS) problem.
Message-passing algorithms stand out as a powerful tool for efficient problem solving within a Bayesian framework.
The performance of message passing relies heavily on the accuracy of the chosen priors, especially the prior of the wireless channel.
Existing works on massive connectivity either assume a statistical channel model, such as Rayleigh fading \cite{chenzhilin2018tsp, liangliu2018tsp, ahn2019ep, guanyongliang2019dnn, chenxiaoming2020modeldriven} and Rician fading \cite{rician2020iotj}, or exploit channel sparsity in certain discrete transform domains, such as the discrete Fourier transform (DFT) angular basis \cite{gaozhen2020tsp, xiaodongwang2022twc}.
The former approach may be overly simplistic to model the multipath component, while the latter suffers from significant energy leakage caused by basis mismatch with the predefined codebook.
The oversimplification or misspecification of the channel prior can result in substantial errors in channel estimation, which consequently degrades the accuracy of activity detection.
Moreover, although not specifically tailored to massive connectivity, the authors in \cite{xiaodongwang2018anm, hangliu2019sr} introduced advanced off-grid and grid-free line spectral estimation techniques for super-resolution channel estimation.
Nevertheless, these methods continue to face challenges in resolving sub-paths within clustered multipath components and often entail prohibitive computational complexity.

In light of the limitations of model-based approaches in accurately capturing the real wireless propagation environments, deep learning offers a powerful alternative by learning the channel distribution directly from data.
Recently, score-based generative models \cite{song2019generative, song2021sde} (a.k.a. diffusion models \cite{ddpm2020nips}) have made remarkable breakthroughs in generating high-quality samples from learned prior distributions, revolutionizing fields such as image synthesis, video generation, and molecular conformation.
These developments highlight their strong ability to model the intricate structures of high-dimensional data distributions.
In particular, score-based generative models learn the prior by matching the gradient of the log density of the data distribution, referred to as the score function.
Apart from generative tasks, such learned priors can also be leveraged to solve CS problems by sampling from the posterior distribution, known as diffusion posterior sampling (DPS) \cite{ddrm2022nips, dps2023iclr, pseudoinverse2023iclr, filtering2024iclr, xue2025rmp, wang2025traversing}.
This approach makes use of the posterior score, which can be obtained by combining the learned prior score with the likelihood term from the measurement model using Bayes' theorem. 
Yet, DPS typically requires traversing hundreds or even thousands of reverse diffusion steps, leading to prohibitively high computational cost.

To speed up convergence, the authors in \cite{cai2025spawc, cai2025score_journal} proposed the integration of score-based generative models into message-passing algorithms for solving inverse problems.
Specifically, a score-based turbo message passing (STMP) framework was developed for compressive image recovery, in which the score network is integrated into Turbo-CS \cite{ma2015turbo_cs} as a plug-and-play denoiser.
This framework inherits the rapid convergence of message passing, typically achieved within $10$ iterations, with each iteration requiring only a single-step denoising operation.
Compared to DPS \cite{ddrm2022nips, dps2023iclr, pseudoinverse2023iclr, filtering2024iclr, xue2025rmp, wang2025traversing}, STMP eliminates the need for reverse diffusion sampling, and thus reduces the computational complexity by orders of magnitude.
This motivates us to explore the feasibility of solving joint activity detection and channel estimation (JADCE) by integrating an accurate channel prior learned by score-based generative networks into message passing.

\subsection{Contributions}
In this paper, we consider a multiple-input multiple-output orthogonal frequency division multiplexing (MIMO-OFDM) grant-free random access network, where a massive number of devices sporadically access the uplink channel.
We aim to develop an efficient message-passing algorithm that incorporates a plug-and-play score-based generative channel prior to approach the performance limit of JADCE.
The extension of STMP \cite{cai2025spawc, cai2025score_journal} for solving the JADCE problem is highly non-trivial, as the sporadic access of devices introduces additional activity-induced sparsity, which is not captured by the channel prior score network.
To address this challenge, we propose a novel turbo message passing (TMP) framework that decouples the denoising of the channel and the device activity.
This stands in contrast to prior works \cite{chenzhilin2018tsp, liangliu2018tsp, gaozhen2020tsp, jiang2022turbo_mp}, which treat the channel and its activity indicator as a whole.
Moreover, unlike the separable denoisers in \cite{chenzhilin2018tsp, liangliu2018tsp, gaozhen2020tsp, jiang2022turbo_mp} that operate element-wise for channel denoising, our TMP framework models the spatial-frequency channel matrix as a super node in the factor graph,
enabling the incorporation of score-based generative models as non-separable denoisers for more effective exploitation of channel correlations.
The score network enables single-step minimum mean-squared error (MMSE) denoising at each message-passing iteration, supported by the well-established Tweedie's formula \cite{robbins1992empirical, efron2011tweedie}.
Additionally, we employ a second-order score network that outputs the Hessian diagonals of the log density to compute the variance of the denoiser.
The plug-in implementation of the first- and second-order score networks yields the STMP algorithm tailored for JADCE, hence the name STMP-JADCE.
The two score networks can be trained based on the denoising score matching technique \cite{vincent2011connection, song2019generative, lu2022maximum}.
The main contributions of this paper are summarized as follows.
\begin{itemize}
	\item We devise a novel TMP framework for JADCE that iterates between a linear estimation module and a denoising module.
	The key innovation lies in the denoising module, which features non-separable channel denoisers and decoupled treatment of channel and activity estimates.
	\item We develop the STMP-JADCE algorithm that seamlessly integrates score-based generative channel priors into the TMP framework.
	As a model-based deep learning solution, STMP-JADCE generalizes well across different device activity levels, pilot patterns, and signal-to-noise ratios (SNRs) without requiring retraining the score networks.
	\item Experimental results show that our method significantly outperforms the existing message-passing algorithms with a comparable or even faster convergence speed.
	Remarkably, our method consistently delivers over a $5$ dB normalized mean-squared error (NMSE) gain in channel estimation on the clustered delay line (CDL) channel models specified by the 3GPP standards \cite{3GPP_TR_38_901}.
\end{itemize}
Beyond the algorithmic advances, this work underscores a conceptual shift in Bayesian inference for massive connectivity: from traditional model-based methods relying on hand-crafted statistical priors (e.g., Rayleigh or Rician fading), to data-driven approaches that directly learn channel distributions from data via score-based generative modeling.
This philosophical transition is central to our contribution, as it enables Bayesian inference to harness expressive learned priors without sacrificing the efficiency of message passing.

\subsection{Related Works}
\subsubsection{Message-Passing Algorithms for JADCE}
It was shown in \cite{chenzhilin2018tsp} that JADCE with single- and multi-antenna BSs corresponds to CS in single measurement vector (SMV) and multiple measurement vector (MMV) setups, respectively.
The authors in \cite{chenzhilin2018tsp} employed the approximate message passing (AMP) to solve both problems, in which the Rayleigh channel assumption is adopted to derive the MMSE denoiser in closed form.
Interestingly, a follow-up work \cite{liangliu2018tsp} revealed that the activity detection error by AMP is asymptotically zero as the number of BS antennas approaches infinity.
Apart from \cite{chenzhilin2018tsp, liangliu2018tsp}, the same Rayleigh channel model was also employed in \cite{ahn2019ep, guanyongliang2019dnn, chenxiaoming2020modeldriven} to facilitate message calculation and analysis.
Considering that the line-of-sight (LoS) component in wave propagation is non-negligible in low Earth orbit (LEO) satellite networks, the authors in \cite{rician2020iotj} proposed a Bernoulli-Rician message-passing algorithm tailored for LEO satellite-enabled IoT applications with Rician channel modeling.
The authors in \cite{cir2017spawc} proposed a JADCE scheme exploiting the sparsity of the time-domain channel impulse response (CIR).
However, an energy leakage problem arises from the inverse discrete Fourier transform (IDFT) to obtain the time-domain channel, which undermines its sparsity level.
Considering frequency-domain channel estimation, the work in \cite{gaozhen2020tsp} developed a turbo-type algorithm that exploits the common angular sparsity over OFDM subcarriers.
To reduce the pilot overhead, the authors in \cite{jiang2022turbo_mp} constructed a block-wise linear model in the frequency domain, requiring the estimation of only the mean and slope in each sub-block.
In the above work, the channel modeling and assigned prior distribution may fail to accurately capture the intricacies of the high-dimensional channel, leading to sub-optimal performance.

\subsubsection{Model-Based and Model-Free Deep Learning for JADCE}
It is well-known that message-passing algorithms can be unfolded into a layer-wise structure analogous to a neural network (NN) \cite{hehengtao2018deep_unfolding}.
By incorporating additional learnable parameters, the unfolded message passing networks can be trained using pairs of the channel samples, device activity realizations, and observations.
It is shown that the deep unfolding technique can mitigate the potential message correlation issue on dense graphs and speed up convergence \cite{guanyongliang2019dnn, chenxiaoming2020modeldriven, xiaodongwang2022twc}.
Nevertheless, deep unfolding, as a model-based deep learning approach, still requires prior knowledge of the wireless channel distribution.
For example, the on-grid millimeter-wave parametric channel is adopted in \cite{xiaodongwang2022twc}, and the Rayleigh fading is adopted in \cite{guanyongliang2019dnn, chenxiaoming2020modeldriven}.
In contrast, pure data-driven approaches \cite{ahn2022deeplearning, zhuhan2023dl} learn to fit certain input-output relationships without the need of explicit mathematical modeling.
In \cite{ahn2022deeplearning}, the authors proposed the use of two long short-term memory (LSTM) networks to learn the mappings from the received signals to the indices of active devices and to the channel coefficients.
However, the existing deep learning-based solutions, whether model-based or model-free, are trained in a supervised manner using predefined active probabilities, specific pilot patterns, and fixed SNRs, which may not generalize effectively to diverse scenarios.

\subsubsection{Score-Based Generative Models for Channel Estimation}
Recent studies \cite{mimoscore2023twc, sure2023asilomar, samsung2024iccwksp, highorderlangevin2024tsp, jcedd2024icassp, fesl2024wcl, jinshi2025twc} have applied DPS \cite{ddrm2022nips, dps2023iclr, pseudoinverse2023iclr, filtering2024iclr, xue2025rmp, wang2025traversing} to solve channel estimation problems by sampling from the posterior channel distribution conditioned on the received pilots. 
Since the pilots (measurement matrix) are solely tied to the likelihood, different pilot designs can be implemented without necessitating retraining of the prior score network. 
As mentioned before, these approaches require hundreds or even thousands of iterations, resulting in unaffordable computational cost. 
Although some initial attempts have explored higher-order annealed Langevin diffusion \cite{highorderlangevin2024tsp} and variational inference \cite{vi2025tcom} to accelerate convergence, the required steps are only reduced by a small margin. % small
Moreover, the authors in \cite{wentaoyu2024jstsp} investigated the integration of score networks into orthogonal AMP (OAMP) \cite{oamp2017access} for channel estimation in near-field holographic MIMO systems.

\subsection{Organization and Notations}
The remainder of this paper is organized as follows.
Section \ref{sec:system_model} presents the system model, formulates the JADCE problem, and outlines the inherent limitations of traditional message-passing algorithms for problem-solving.
Section \ref{sec:tmp_framework} develops a TMP framework that supports the integration of score-based generative priors.
Section \ref{sec:loss_functions} reveals the relationship between the score function and MMSE denoising, and then specifies the training objectives for the score networks.
In Section \ref{sec:stmp}, we present the STMP-JADCE algorithm and related engineering tricks.
In Section \ref{sec:simulations}, we provide extensive simulation results to validate the efficiency and effectiveness of the proposed method.
Finally, we conclude this paper in Section \ref{sec:conclusions}.

\textit{Notations:}
Lower-case letters are used to denote scalars.
Vectors and matrices are denoted by lower-case and upper-case boldface letters, respectively.
We use $\mbf{A}^{\sf T}$, $\mbf{A}^{\sf H}$, and $\mbf{A}^{-1}$ to denote the transpose, conjugate transpose, and inverse of matrix $\mbf{A}$, respectively.
We use the symbol $\propto$ to indicate equality up to a constant multiplicative factor.
We use $\diag\{ \cdot \}$, $\mathrm{tr} ( \cdot )$, $\mathbb{E} [ \cdot ]$, $\mathrm{Var} [\cdot]$, and $\mathrm{Cov} [\cdot]$ to represent the diagonal operator, the trace of a square matrix, the expectation, the variance, and the covariance matrix, respectively.
The probability density function (pdf) of a circularly-symmetric complex Gaussian (CSCG) random variable $x$ with mean $m$ and variance $v$ is denoted by $\mathcal{CN}(x; m, v)$.
The pdf of a CSCG random vector $\mbf{x}$ with mean $\mbf{m}$ and covariance matrix $\mbf{V}$ is denoted by $\mathcal{CN}(\mbf{x}; \mbf{m}, \mbf{V})$.

\begin{figure}
	[t]
	\centering
	%	\vspace{-1em}
	\includegraphics[width=.95\columnwidth]{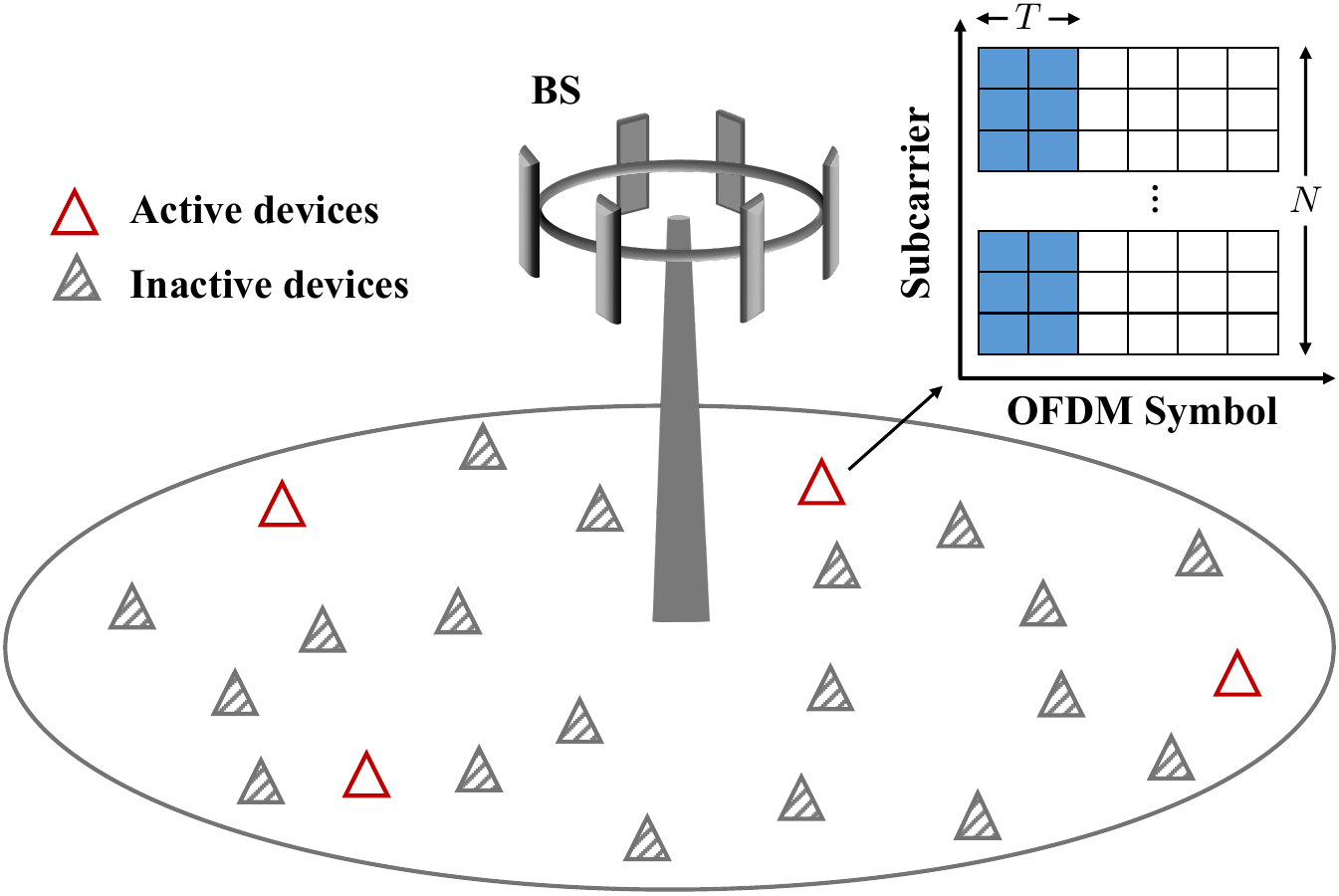}
	%	\vspace{-.5em}
	\caption{System model of the MIMO-OFDM grant-free random access network.}
	\label{fig:system_model}
	%	\vspace{-.5em}
\end{figure}

\section{System Model and Problem Statement} \label{sec:system_model}
\subsection{System Model}
Consider the uplink transmission of massive devices in an IoT network, as illustrated in Fig. \ref{fig:system_model}.
The system operates under a grant-free random access scheme to support the sporadic connectivity of $K\gg 1$ devices.
The BS is equipped with $M$ antennas, while each IoT device is equipped with a single antenna.
A total of $N$ adjacent OFDM subcarriers are shared among the devices for pilot transmission.
We adopt a block-fading model, i.e., all the channels remain quasi-static within a transmission block.
Moreover, we assume that the IoT devices are synchronized, and each device determines whether to access the channel in an independent and identically distributed (i.i.d.) manner.
As a result, only a (small) subset of the devices is active in any given transmission block.
We use an indicator function $\alpha_k$ to model the device activity as  
\begin{align}
	\alpha_k = \begin{cases}
		1, & \text{if the $k$-th device is active,} \\
		0, &\text{otherwise.}
	\end{cases}
\end{align}
Let $\lambda$ represent the probability of a device being active, i.e., $p(\alpha_k=1) = \lambda$, where $\lambda \ll 1$.

Denote by $h_{knm} \in \mathbb{C}$ the channel frequency response on the $n$-th subcarrier from the $k$-th device to the $m$-th BS antenna.
Then, the channel frequency response from the $k$-th device to the BS can be expressed in a matrix form as
\begin{align}
	\mbf{H}_k = \begin{bmatrix}
		h_{k11} & h_{k12} & \dots & h_{k1M} \\
		h_{k21} & h_{k22} & \dots & h_{k2M} \\
		\vdots & \vdots & \ddots & \vdots \\
		h_{kN1} & h_{kN2} &  \dots & h_{kNM} 
	\end{bmatrix} \in \mathbb{C}^{N \times M}.
\end{align}
Assume that a total of $T$ OFDM symbols are allocated for pilot transmission.
Let $q_{ktn} \in \mathbb{C}$ be the pilot symbol transmitted by the $k$-th device during the $t$-th OFDM symbol on the $n$-th subcarrier.
We construct a diagonal pilot matrix as $\mbf{Q}_{kt} \triangleq \diag \left\{q_{kt1}, \dots, q_{ktN}\right\} \in \mathbb{C}^{N \times N}$, $\forall k, t$.
Assume that the cyclic prefix (CP) length exceeds the maximum channel delay spread.
After removing the CP and applying the DFT, the received signal $\mbf{Y}_t \in \mathbb{C}^{N \times M}$ during the $t$-th OFDM symbol is given by
\begin{align}
	\mathbf{Y}_t = \sum_{k=1}^K \alpha_k \mbf{Q}_{kt} \mathbf{H}_k + \mathbf{N}_t, \label{eqn:sys_model_H}
\end{align}
where $\mathbf{N}_t \in \mathbb{C}^{N\times M}$ is an AWGN matrix with the elements i.i.d. drawn from $\mathcal{CN}(0, \delta_0^2)$.
Let $\mbf{X}_k \triangleq \alpha_k \mbf{H}_k$.
Define $\mbf{Y} \triangleq  \left[\mbf{Y}_1^{\sf H}, \dots, \mbf{Y}_T^{\sf H}\right]^{\sf H} \in \mathbb{C}^{NT \times M}$,
$\mbf{Q}_k \triangleq \left[\mbf{Q}_{k1}^{\sf H}, \dots, \mbf{Q}_{kT}^{\sf H}\right]^{\sf H} \in \mathbb{C}^{NT \times N}$, and
$\mbf{N} \triangleq \left[\mbf{N}_1^{\sf H}, \dots, \mbf{N}_T^{\sf H}\right]^{\sf H} \in \mathbb{C}^{NT \times M}$.
We express the received signal $\mbf{Y}$ over the $T$ OFDM symbols in a more compact form as
\begin{align}
	\mbf{Y} = \sum_{k=1}^K \mbf{Q}_k \mbf{X}_k + \mbf{N}
	= \mbf{Q} \mbf{X} + \mbf{N},
\end{align}
where $\mbf{Q} \triangleq \left[\mbf{Q}_1, \dots, \mbf{Q}_K\right] \in \mathbb{C}^{NT\times NK}$, and
$\mbf{X} \triangleq \left[\mbf{X}_1^{\sf H}, \dots, \mbf{X}_K^{\sf H}\right]^{\sf H} \in \mathbb{C}^{NK \times M}$.
For simplicity, we impose the condition that $\mbf{Q}$ is partial orthogonal, specifically requiring that $\mbf{Q}\mbf{Q}^{\sf H} = KP \mbf{I}_{NT}$, where each device has the same transmit power budget $P$.
The design of partial orthogonal $\mbf{Q}$ is deferred to Section \ref{sec:pilot_design}.
Due to the sporadic access of the devices, matrix $\mathbf{X}$ exhibits a block-sparsity structure.
Specifically, if the $k$-th device is inactive, we have $\mathbf{X}_k = \alpha_k \mathbf{H}_k =\mathbf{0}$.
Frequently used notations in this paper are summarized in Table~\ref{table:notations}.

\begin{table}[t]
	\caption{Summary of Notations}
	\label{table:notations}
	\centering
	\begin{tabular}{rl}
		\toprule
		Notation & Description \\
		\midrule
		$M$ & Number of BS antennas  \\
		$N$ & Number of OFDM subcarriers   \\
		$K$ & Number of sporadic access devices  \\
		$T$ & Number of OFDM symbols \\
		$P$ & Transmit power budget of each device \\
		$\lambda$ & Prior probability that a device is active  \\
		$\alpha_k$ &  Indicator of whether device $k$ is active \\
		$\mbf{H}_k$ & Channel matrix from device $k$ to the BS \\
		$\mbf{X}_k$ & Effective channel matrix $\mbf{X}_k \triangleq \alpha_k \mbf{H}_k$ \\
		$\mbf{h}_{km}$ ($\mbf{x}_{km}$) & The $m$-th column of $\mbf{H}_k$ ($\mbf{X}_k$) \\
		$\mbf{X}$ & $\mbf{X} \triangleq \left[\mbf{X}_1^{\sf H}, \dots, \mbf{X}_K^{\sf H}\right]^{\sf H}$ \\
		$\mbf{x}_m$ & The $m$-th column of $\mbf{X}$ \\
		$\mbf{Q}$ & Concatenated pilot symbol matrix \\
		$\mbf{Y}$ & Concatenated received signal matrix \\
		$\mbf{N}$ & Concatenated AWGN matrix \\
		%		\midrule
		%
		$\mbf{x}_{m,\mathsf{A}}^\mathtt{pri}$, $v_{m,\mathsf{A}}^\mathtt{pri}$ & Prior mean and variance of $\mbf{x}_m$ in module A \\
		$\mbf{x}_{m,\mathsf{A}}^\mathtt{post}$, $v_{m,\mathsf{A}}^\mathtt{post}$ & Posterior mean and variance of $\mbf{x}_m$ in module A \\
		$\mbf{x}_{m,\mathsf{A}}^\mathtt{ext}$, $v_{m,\mathsf{A}}^\mathtt{ext}$ & Extrinsic mean and variance of $\mbf{x}_m$ in module A \\
		$\mbf{x}_{km,\mathsf{B}}^\mathtt{pri}$, $v_{m,\mathsf{B}}^\mathtt{pri}$ & Prior mean and variance of $\mbf{x}_{km}$ in module B \\
		$\mbf{x}_{km,\mathsf{B}}^\mathtt{post}$, $v_{m,\mathsf{B}}^\mathtt{post}$ & Posterior mean and variance of $\mbf{x}_{km}$ in module B \\
		$\mbf{x}_{km,\mathsf{B}}^\mathtt{ext}$, $v_{m,\mathsf{B}}^\mathtt{ext}$ & Extrinsic mean and variance of $\mbf{x}_{km}$ in module B \\
		$\mbf{h}_{km}^\mathtt{pri}$, $\tau_{m}^\mathtt{pri}$ & Prior mean and variance of $\mbf{h}_{km}$ in module B \\
		$\mbf{h}_{km}^\mathtt{post}$, $\tau_{m}^\mathtt{post}$ & Posterior mean and variance of $\mbf{h}_{km}$ in module B \\
		$\mbf{h}_{km}^\mathtt{ext}$, $\tau_{m}^\mathtt{ext}$ & Extrinsic mean and variance of $\mbf{h}_{km}$ in module B \\
		$\lambda_k^\mathtt{post}$ & Posterior estimate of the active probability of device $k$ \\
		\bottomrule
	\end{tabular}
\end{table}

\subsection{Problem Statement}
In this paper, we aim to simultaneously detect the active devices and estimate the corresponding channels based on the observation $\mathbf{Y}$ under a Bayesian inference framework.
Exact Bayesian estimators, e.g., the MMSE estimator, are generally intractable due to the high-dimensional integrals in the marginalization of the joint posterior distribution $p\big(\mathbf{X}, \{\mathbf{H}_k\}_{k=1}^K, \{\alpha_k\}_{k=1}^K|\mathbf{Y}\big)$.

\begin{figure}
	[t]
	\centering
	%	\vspace{-1em}
	\includegraphics[width=1\columnwidth]{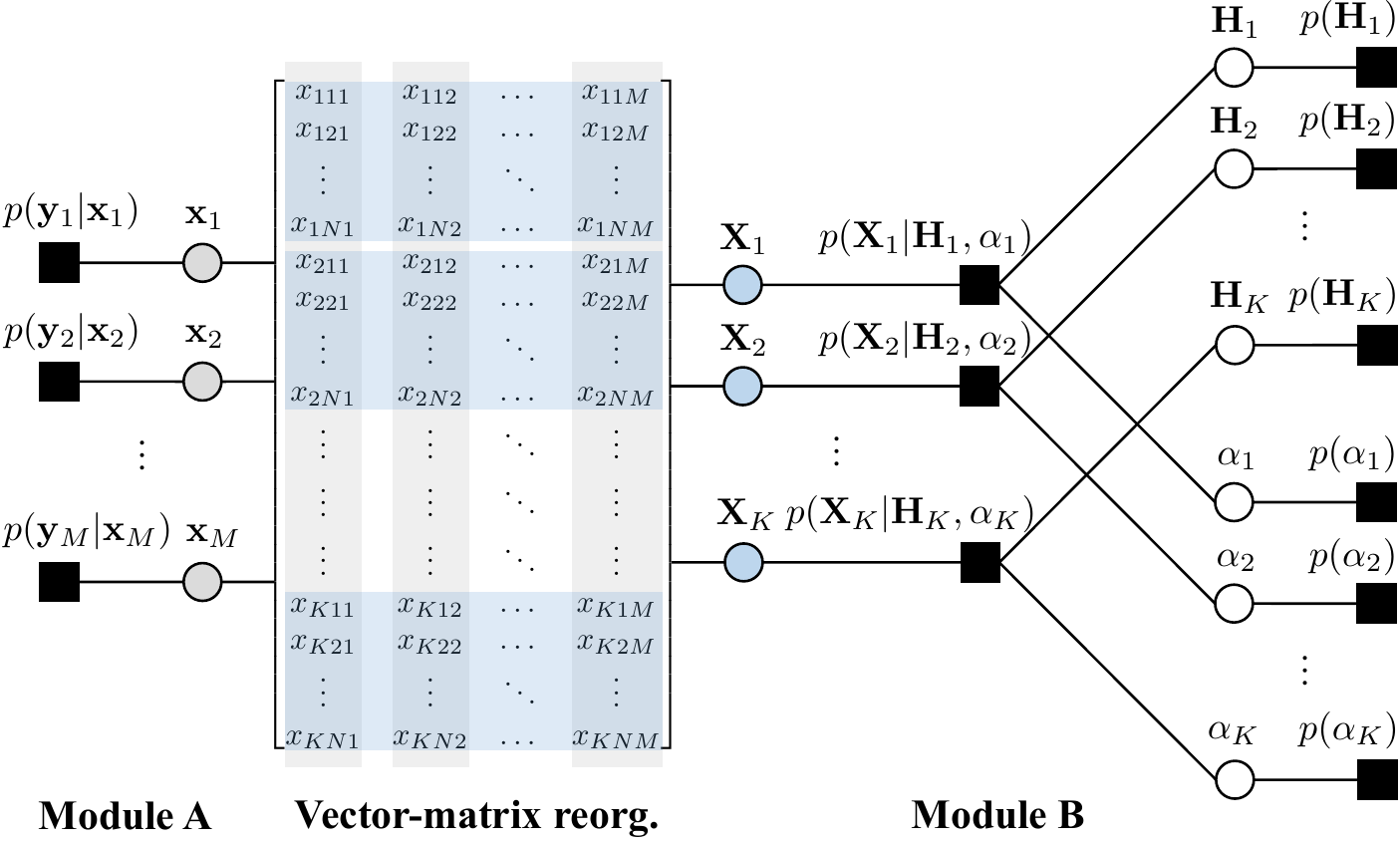}
	%		\vspace{-1.5em}
	\caption{Factor graph representation of the joint posterior distribution $p(\mathbf{X}, \{\mathbf{H}_k\}_{k=1}^K, \{\alpha_k\}_{k=1}^K|\mathbf{Y})$.
		The variables are represented by the ``variable nodes'' that appear as circles, while the distributions are represented by the ``factor nodes'' that appear as black filled squares.
		We use different colors to differentiate $\mbf{x}_m$ and $\mbf{X}_k$ in vector-matrix reorganization.}
	\label{fig:factor_graph}
	%		\vspace{-.5em}
\end{figure}

Alternatively, message-passing algorithms \cite{liangliu2018tsp, chenzhilin2018tsp, jiang2022turbo_mp, bianxinyu2023joint} provide a low-complexity iterative approach for approximately calculating the MMSE estimator.
In message passing, we factorize the joint posterior distribution according to Baye's theorem and the conditional independence relationships among the variables, expressed as
\begin{align}
	& p\big(\mathbf{X}, \{\mathbf{H}_k\}_{k=1}^K, \{\alpha_k\}_{k=1}^K|\mathbf{Y}\big)\nonumber \\
	&\propto \prod_{m=1}^M p(\mathbf{y}_m |\mathbf{x}_m)
	\prod_{k=1}^K p(\mathbf{X}_k |\mbf{H}_k, \alpha_k)
	 p(\mbf{H}_k)p(\alpha_k) , \label{posterior_distribution}
\end{align}
where $\mbf{x}_m \in \mathbb{C}^{NK}$ and $\mbf{y}_m \in \mathbb{C}^{NT}$ denote the $m$-th column of $\mbf{X} = \left[\mbf{x}_1, \dots, \mbf{x}_M\right]$ and $\mbf{Y}= \left[\mbf{y}_1, \dots, \mbf{y}_M\right]$, respectively.
A factor graph is then utilized to represent the joint distribution in \eqref{posterior_distribution}, as illustrated in Fig. \ref{fig:factor_graph}. 
The distributions of each factor node and their abbreviations are summarized in Table~\ref{table:factor_nodes}.
The main difficulty in implementing message passing on the factor graph lies in the message calculation associated with the factor node $p(\mbf{H}_k)$, detailed as follows.
\begin{itemize}
	\item \textbf{Lack of exact characterization of channel distribution:} It is extremely challenging, if not impossible, to obtain an accurate characterization of the channel distribution $p(\mbf{H}_k)$.
	Existing literature resorts to various assumptions, e.g., Rayleigh fading \cite{liangliu2018tsp, chenzhilin2018tsp}, Rician fading \cite{rician2020iotj}, angular-frequency structured sparsity \cite{gaozhen2020tsp}, or frequency-domain block-wise linearity \cite{jiang2022turbo_mp}, to model the wireless channel.
	These assumptions, however, may be mismatched with the ground-truth prior distribution, leading to significant performance degradation and potentially causing the algorithm to diverge.
	\item \textbf{Intractable high-dimensional integrals:} The messages associated with the factor node $p(\mbf{H}_k)$ can only be derived under specific, simple, and easily computable priors.
	For example, the work in \cite{gaozhen2020tsp} adopted a simple i.i.d. Bernoulli-Gaussian distribution to account for the inherent sparsity in the angular domain.
	Generally speaking, even if the ground-truth channel distribution becomes analytically available, calculating the messages for the factor node $p(\mbf{H}_k)$ remains highly challenging  due to the complicated and high-dimensional integrals (as will be discussed in Section \ref{sec:channel_device_denoisers}).
\end{itemize}
To summarize, the message-passing framework faces a dilemma: the adopted channel prior is either too imprecise to ensure good performance, or too complex to facilitate efficient message calculation.
This paper seeks to address the aforementioned challenges by plugging in deep generative priors for efficient message calculation.
We begin by developing a TMP framework for JADCE.
Then, we present a unique interpretation of the belief calculation at the factor node $p(\mathbf{H}_k)$ as an MMSE denoising problem under AWGN observations.
This insight enables the integration of score-based generative networks for provable MMSE denoising, eliminating the need to compute high-dimensional integrals. 

\begin{table}[t]
	\caption{Factor Nodes and Their Abbreviations}
	\label{table:factor_nodes}
	\centering
	\begin{tabular}{ccc}
		\toprule
		Factor node & Abbreviation & Distribution \\
		\midrule
		$p(\mbf{y}_m|\mbf{x}_m)$ & $f_{\mbf{y}_m}$  & $\mathcal{CN}(\mbf{Q}\mbf{x}_m, \delta_0^2\mbf{I}_{NT})$  \\
		$p(\mbf{X}_k|\mbf{H}_k, \alpha_k)$ & $f_{\mbf{X}_k}$  & $\delta (\mbf{X}_k - \alpha_k\mbf{H}_k)$   \\
		$p(\mbf{H}_k)$ & $f_{\mbf{H}_k}$ & -- \\
		$p(\alpha_k)$ & $f_{\alpha_k}$ & $(1-\lambda)\delta (\alpha_k) + \lambda \delta(1-\alpha_k) $ \\
		\bottomrule
	\end{tabular}
\end{table}

\section{Turbo Message Passing Framework} \label{sec:tmp_framework}
In this section, we develop a novel TMP framework based on the combined use of the sum-product rule \cite{sum_product_algo2001tit} and the EP algorithm \cite{EP2001Minka}.
TMP interprets the message updates on the factor graph in Fig. \ref{fig:factor_graph} as iterations between two modules, namely, module A and module B.
Module A is a linear estimator that separately  estimates each column of $\mbf{X}$ (i.e., $\mbf{x}_m$) exploiting the likelihood $p(\mbf{y}_m|\mbf{x}_m)$ and the messages from module B.
Module B is a denoiser of $\mbf{X}_k$ exploiting the priors $p(\mbf{H}_k)$ and $p(\alpha_k)$, along with the messages from module A.
These estimates iterate between the two modules like turbo decoding, hence the name TMP.
We describe the message calculations in module A and module B in the following subsections.

\subsection{Module A: Linear Estimator} \label{sec:module_A}
Following Turbo-CS \cite{ma2015turbo_cs}, we start by initializing the message from the variable node $\mbf{x}_m$ to the factor node $p(\mbf{y}_m|\mbf{x}_m)$ as $\mathcal{M}_{\mathbf{x}_m \rightarrow f_{\mbf{y}_m}}(\mathbf{x}_m) = \mathcal{CN} (\mathbf{x}_m; \mathbf{x}_{m,\mathsf{A}}^\mathtt{pri}, v_{m,\mathsf{A}}^{\mathtt{pri}}\mathbf{I}_{NK})$.
The belief of $\mbf{x}_m$ on the factor node $p(\mbf{y}_m|\mbf{x}_m)$ is expressed as
\begin{align} \label{eqn:belief_x_m}
	\mathcal{M}_{f_{\mbf{y}_m}}(\mathbf{x}_m) \propto p(\mathbf{y}_m|\mathbf{x}_m) \mathcal{M}_{\mathbf{x}_m \rightarrow f_{\mbf{y}_m}}(\mathbf{x}_m).
\end{align}
In \eqref{eqn:belief_x_m}, the belief $\mathcal{M}_{f_{\mbf{y}_m}}(\mathbf{x}_m)$ is proportional to the product of two Gaussian pdfs, which is also a Gaussian pdf.
The mean and covariance can be obtained by completing the square in the exponent of the Gaussian product with respect to $\mathbf{x}_m$.
We further approximate the covariance matrix using a shared scalar variance \cite{ma2015turbo_cs, vamp2019tit}, resulting in the belief of the form $\mathcal{M}_{f_{\mbf{y}_m}}(\mathbf{x}_m) \approx \mathcal{CN}(\mbf{x}_m; \mbf{x}_{m, \mathsf{A}}^\mathtt{post}, v_{m, \mathsf{A}}^\mathtt{post} \mbf{I}_{NK})$, where
\begin{align}
	\mbf{x}_{m,\mathsf{A}}^\mathtt{post} &= \mbf{x}_{m,\mathsf{A}}^\mathtt{pri} + \frac{v_{m,\mathsf{A}}^\mathtt{pri}}{KPv_{m,\mathsf{A}}^\mathtt{pri}+\delta_0^2} \mbf{Q}^{\sf H} \left(\mbf{y}_m-\mbf{Q} \mbf{x}_{m,\mathsf{A}}^\mathtt{pri} \right), \label{eqn:X_post_A} \\
	v_{m, \mathsf{A}}^\mathtt{post} &=  v_{m, \mathsf{A}}^\mathtt{pri} - \frac{TP (v_{m, \mathsf{A}}^\mathtt{pri})^2}{KP  v_{m, \mathsf{A}}^\mathtt{pri}+\delta_0^2}. \label{eqn:v_post_A}
\end{align}
Eqns. \eqref{eqn:X_post_A} and \eqref{eqn:v_post_A} are respectively the posterior mean and variance from the linear MMSE (LMMSE) estimator of $\mbf{x}_m$, given the prior mean $\mbf{x}_{m,\mathsf{A}}^\mathtt{pri}$ and variance $v_{m,\mathsf{A}}^\mathtt{pri}$.

The message from the factor node $p(\mbf{y}_m|\mbf{x}_m)$ to the variable node $\mbf{x}_m$, also known as the extrinsic message, is calculated based on the EP algorithm \cite{EP2001Minka} as
\begin{align}
	\mathcal{M}_{f_{\mbf{y}_m}\rightarrow \mbf{x}_m}(\mbf{x}_m) \propto \frac{\mathcal{M}_{f_{\mbf{y}_m}}(\mathbf{x}_m)}{\mathcal{M}_{\mathbf{x}_m \rightarrow f_{\mbf{y}_m}}(\mathbf{x}_m)} . \label{moduleA_ext}
\end{align}
From \eqref{moduleA_ext}, the extrinsic message is proportional to the ratio of two Gaussian pdfs, which is another Gaussian pdf, denoted as $\mathcal{M}_{f_{\mbf{y}_m}\rightarrow \mbf{x}_m}(\mbf{x}_m) = \mathcal{CN} (\mathbf{x}_m; \mathbf{x}_{m,\mathsf{A}}^\mathtt{ext}, v_{m,\mathsf{A}}^{\mathtt{ext}}\mathbf{I}_{NK})$.
The variance $v_{m,\mathsf{A}}^{\mathtt{ext}}$ and mean $\mathbf{x}_{m,\mathsf{A}}^\mathtt{ext}$ are respectively given by
\begin{align}
	v_{m,\mathsf{A}}^{\mathtt{ext}} &= \left(\frac{1}{v_{m,\mathsf{A}}^{\mathtt{post}}} - \frac{1}{v_{m,\mathsf{A}}^{\mathtt{pri}}}\right)^{-1}, \label{eqn:v_ext_A} \\
	\mathbf{x}_{m,\mathsf{A}}^\mathtt{ext} &= v_{m,\mathsf{A}}^{\mathtt{ext}} \left(\frac{\mathbf{x}_{m,\mathsf{A}}^\mathtt{post}}{v_{m,\mathsf{A}}^{\mathtt{post}}} - \frac{\mathbf{x}_{m,\mathsf{A}}^\mathtt{pri}}{v_{m,\mathsf{A}}^{\mathtt{pri}}}\right). \label{eqn:X_ext_A}
\end{align}
The message $\mathcal{M}_{f_{\mbf{y}_m}\rightarrow \mbf{x}_m}(\mbf{x}_m)$ flows rightward through the factor graph unchanged, serving as the input to module B: $\mathbf{x}_{m,\mathsf{B}}^\mathtt{pri} = \mathbf{x}_{m,\mathsf{A}}^\mathtt{ext}$, $v_{m,\mathsf{B}}^{\mathtt{pri}} = v_{m,\mathsf{A}}^{\mathtt{ext}}$.

\subsection{Module B: Channel and Device Activity Denoisers} \label{sec:channel_device_denoisers}
As shown in Fig. \ref{fig:factor_graph}, the denoiser of $\mbf{X}_k = \alpha_k \mbf{H}_k$ can be further divided into two sub-modules: the denoiser of $\mbf{H}_k$ and the denoiser of $\alpha_k$.
The outputs from these two sub-modules are then combined at the factor node $p(\mbf{X}_k|\mbf{H}_k, \alpha_k)$ to produce the denoised $\mbf{X}_k$. 
In this subsection, we formally establish the equivalence between belief calculation at the factor node $p(\mbf{H}_k)$ and MMSE denoising of $\mbf{H}_k$ under AWGN observations.
This equivalence plays a crucial role in the integration of score-based generative priors for efficient belief calculation.

Firstly, we derive the message from the variable node $\mbf{X}_k$ to the factor node $p(\mbf{X}_k|\mbf{H}_k,\alpha_k)$, denoted as $\mathcal{M}_{\mbf{X}_k \rightarrow f_{\mbf{X}_k}} (\mbf{X}_k)$, by examining the relationships in vector-matrix reorganization.
We marginalize the message $\mathcal{M}_{f_{\mbf{y}_m}\rightarrow \mbf{x}_m}(\mbf{x}_m) = \mathcal{CN} \big(\mathbf{x}_m; \mathbf{x}_{m,\mathsf{B}}^\mathtt{pri}, v_{m,\mathsf{B}}^{\mathtt{pri}}\mathbf{I}_{NK}\big)$ from module A to obtain
\begin{align}
	\mathcal{M}_{f_{\mbf{y}_m}\rightarrow \mbf{x}_{km}}(\mbf{x}_{km}) = \mathcal{CN} \big(\mbf{x}_{km}; \mbf{x}_{km,\mathsf{B}}^\mathtt{pri}, v_{m,\mathsf{B}}^{\mathtt{pri}}\mbf{I}_{N}\big) ,\label{column_independence}
\end{align}
where $\mbf{x}_{km} \triangleq \left[x_{k1m}, \dots, x_{kNm}\right]^{\sf T} \in \mathbb{C}^N$ and
$\mbf{x}_{km,\mathsf{B}}^\mathtt{pri} \triangleq \big[x_{k1m,\mathsf{B}}^\mathtt{pri}, \dots, x_{kNm,\mathsf{B}}^\mathtt{pri}\big]^{\sf T} \in \mathbb{C}^N$ represent the $k$-th blocks in $\mbf{x}_m$ and $\mathbf{x}_{m,\mathsf{B}}^\mathtt{pri}$, respectively,
i.e., $\mbf{x}_m = \left[\mbf{x}_{1m}^{\sf H}, \dots, \mbf{x}_{Km}^{\sf H}\right]^{\sf H}$ and $\mathbf{x}_{m,\mathsf{B}}^\mathtt{pri} = \big[(\mathbf{x}_{1m,\mathsf{B}}^\mathtt{pri})^{\sf H}, \dots, (\mathbf{x}_{Km,\mathsf{B}}^\mathtt{pri})^{\sf H}\big]^{\sf H}$. 
After vector-matrix reorganization,
the message from the variable node $\mbf{X}_k$ to the factor node $p(\mbf{X}_k|\mbf{H}_k,\alpha_k)$ manifests  as
\begin{align}
	\mathcal{M}_{\mbf{X}_k \rightarrow f_{\mbf{X}_k}} (\mbf{X}_k) &= \prod_{m} \mathcal{M}_{f_{\mbf{y}_m}\rightarrow \mbf{x}_{km}}(\mbf{x}_{km}) \label{column_independent}  \\
	&= \prod_{m} \mathcal{CN} \big(\mbf{x}_{km}; \mbf{x}_{km, \mathsf{B}}^\mathtt{pri}, v_{m, \mathsf{B}}^\mathtt{pri} \mbf{I}_N\big),
\end{align}
where \eqref{column_independent} arises from the fact that the messages $\mathcal{M}_{f_{\mbf{y}_m}\rightarrow \mbf{x}_{km}}(\mbf{x}_{km})$  are independent for different values of $m$.

\subsubsection{Denoiser of $\mbf{H}_k$}
The message from the factor node $p(\mbf{X}_k|\mbf{H}_k,\alpha_k)$ to the variable node $\mbf{H}_k$ is computed following the sum-product rule \cite{sum_product_algo2001tit}:
\begin{align}
	&\mathcal{M}_{f_{\mbf{X}_k} \rightarrow \mbf{H}_k}(\mbf{H}_k) \nonumber \\
	&\propto \int_{\mbf{X}_k, \alpha_k} \delta (\mbf{X}_k -\alpha_k \mbf{H}_k) 
	\mathcal{M}_{\mbf{X}_k \rightarrow f_{\mbf{X}_k}} (\mbf{X}_k)
	p(\alpha_k)
	 \nonumber \\
	&\propto (1-\lambda) \int_{\mbf{X}_k} \delta (\mbf{X}_k) \prod_{m} \mathcal{CN} \big(\mbf{x}_{km}; \mbf{x}_{km, \mathsf{B}}^\mathtt{pri}, v_{m, \mathsf{B}}^\mathtt{pri} \mbf{I}_N\big) \nonumber\\
	&~~~\; + \lambda \int_{\mbf{X}_k} \delta(\mbf{X}_k - \mbf{H}_k)  \prod_{m} \mathcal{CN} \big(\mbf{x}_{km}; \mbf{x}_{km, \mathsf{B}}^\mathtt{pri}, v_{m, \mathsf{B}}^\mathtt{pri} \mbf{I}_N\big) \nonumber \\
	&\propto (1-\lambda) \prod_{m} \mathcal{CN} \big(\mbf{0}; \mbf{x}_{km, \mathsf{B}}^\mathtt{pri}, v_{m, \mathsf{B}}^\mathtt{pri} \mbf{I}_N\big) \nonumber \\
	&~~~\; + \lambda \prod_{m} \mathcal{CN} \big(\mbf{h}_{km}; \mbf{x}_{km, \mathsf{B}}^\mathtt{pri}, v_{m, \mathsf{B}}^\mathtt{pri} \mbf{I}_N\big), \label{const_gaussian}
\end{align}
where $\mbf{h}_{km} \triangleq \left[h_{k1m}, \dots, h_{kNm}\right]^{\sf T} \in \mathbb{C}^N$ represents the $m$-th column of $\mbf{H}_k$, i.e., $\mbf{H}_k = \left[\mbf{h}_{k1}, \dots, \mbf{h}_{kM}\right]$.
The first term in \eqref{const_gaussian} corresponds to the case that the $k$-th device is inactive and is a constant with respect to $\mbf{H}_k$.
As a result, $\mathcal{M}_{f_{\mbf{X}_k} \rightarrow \mbf{H}_k}(\mbf{H}_k)$ cannot be properly normalized to a valid pdf when this term is retained. 
To obtain a tractable message for channel denoising, we approximate 
$\mathcal{M}_{f_{\mbf{X}_k} \rightarrow \mbf{H}_k}(\mbf{H}_k)$ by conditioning on the active case and forwarding only the second term to the channel denoiser.
The activity uncertainty will be handled separately through the Bernoulli update in \eqref{eqn:lambda_post}.
The approximated Gaussian message reads
\begin{align}
	\mathcal{M}_{f_{\mbf{X}_k} \rightarrow \mbf{H}_k}(\mbf{H}_k) \approx \prod_{m} \mathcal{CN} \big(\mbf{h}_{km}; \mbf{h}_{km}^\mathtt{pri}, \tau_{m}^\mathtt{pri} \mbf{I}_N\big),
\end{align}
where $\mbf{h}_{km}^\mathtt{pri} = \mbf{x}_{km, \mathsf{B}}^\mathtt{pri}$ and $\tau_{m}^\mathtt{pri} = v_{m, \mathsf{B}}^\mathtt{pri}$.

Based on the update rules of EP \cite{EP2001Minka}, 
the belief of $\mbf{H}_k$ on the factor node $p(\mbf{H}_k)$ is expressed as
\begin{align}
	\mathcal{M}_{f_{\mbf{H}_k}}(\mbf{H}_k) &= \mathrm{proj}\left[p(\mbf{H}_k) \mathcal{M}_{f_{\mbf{X}_k} \rightarrow \mbf{H}_k}(\mbf{H}_k) \right]  \nonumber \\
	&\approx \prod_{m} \mathcal{CN}\big(\mbf{h}_{km}; \mbf{h}_{km}^\mathtt{post}, \tau_{m}^\mathtt{post}\mbf{I}_N\big), \label{v_H_post}
\end{align}
where $\mathrm{proj}[\cdot]$ represents the projection of a probability distribution to a Gaussian density with matched first- and second-order moments, and \eqref{v_H_post} approximates the variance by a set of scalars.
The posterior mean $\mbf{h}_{km}^\mathtt{post}$ and variance $\tau_{m}^\mathtt{post}$ can be calculated as
\begin{align}
	\!\!\mbf{h}_{km}^\mathtt{post} \!&= \frac{1}{Z_k} \int_{\mbf{H}_k} \!\!\mbf{h}_{km} p(\mbf{H}_k) \prod_{m} \mathcal{CN} \big(\mbf{h}_{km}; \mbf{h}_{km}^\mathtt{pri}, \tau_{m}^\mathtt{pri} \mbf{I}_N\big), \label{eqn:post_mean_integral} \\
	\!\!\tau_{m}^\mathtt{post} \!&=  \frac{1}{KN}\sum_{k,n}\frac{1}{Z_k}  \int_{\mbf{H}_k} |h_{knm} - h_{knm}^\mathtt{pri}|^2 p(\mbf{H}_k) \nonumber \\
	&~~~~~~~~~~~~~~~~~~~~~~ \times \prod_{m} \mathcal{CN} \big(\mbf{h}_{km}; \mbf{h}_{km}^\mathtt{pri}, \tau_{m}^\mathtt{pri} \mbf{I}_N\big), \label{eqn:post_var_integral}
\end{align}
where $Z_k \triangleq \int_{\mbf{H}_k} p(\mbf{H}_k) \prod_{m} \mathcal{CN} \big(\mbf{h}_{km}; \mbf{h}_{km}^\mathtt{pri}, \tau_{m}^\mathtt{pri} \mbf{I}_N\big)$ is the normalization constant.

State evolution (SE) analysis \cite{se_analysis2018nips} has shown that the error vector $\mbf{h}_{km}^\mathtt{pri} - \mbf{h}_{km}$ is i.i.d. Gaussian in the large system limit.
In other words, module A decouples the linear mixing model \eqref{eqn:sys_model_H} into a series of independent AWGN observations of the ground truth, i.e.,
\begin{align}
	\mbf{h}_{km}^\mathtt{pri} = \mbf{h}_{km} + \mbf{w}_{km}, ~~ \mbf{w}_{km} \sim \mathcal{CN}\big(\mbf{0}, \tau_{m}^\mathtt{pri} \mbf{I}_N\big). \label{eqn:awgn_observation}
\end{align}
Eqn. \eqref{eqn:awgn_observation} implies the Gaussian transition probability $p(\mbf{h}_{km}^\mathtt{pri}|\mbf{h}_{km}) = \mathcal{CN} \big(\mbf{h}_{km}^\mathtt{pri}; \mbf{h}_{km}, \tau_{m}^\mathtt{pri} \mbf{I}_N\big)$, and therefore,
\begin{align}
	p(\mbf{H}_{k}^\mathtt{pri}|\mbf{H}_k) = \prod_{m} \mathcal{CN} \big(\mbf{h}_{km}^\mathtt{pri}; \mbf{h}_{km}, \tau_{m}^\mathtt{pri} \mbf{I}_N\big),
\end{align}
where we denote $\mbf{H}_{k}^\mathtt{pri} \triangleq\big[\mbf{h}_{k1}^\mathtt{pri}, \dots, \mbf{h}_{kM}^\mathtt{pri}\big]$ for brevity.
By noting that $\prod_{m} \mathcal{CN} \big(\mbf{h}_{km}; \mbf{h}_{km}^\mathtt{pri}, \tau_{m}^\mathtt{pri} \mbf{I}_N\big)$ in \eqref{eqn:post_mean_integral} and \eqref{eqn:post_var_integral} takes the same form as $p(\mbf{H}_{k}^\mathtt{pri}|\mbf{H}_k)$, the belief calculation of $\mbf{H}_k$ can be interpreted as an MMSE denoising problem given the AWGN model \eqref{eqn:awgn_observation}: 
\begin{align}
	\mbf{H}_{k}^\mathtt{post} &= \mathbb{E}\big[\mbf{H}_k|\mbf{H}_{k}^\mathtt{pri}\big], \label{eqn:post_mean}\\
	\tau_{m}^\mathtt{post} &= \frac{1}{KN} \sum_{k,n}\mathrm{Var}\big[h_{knm}|\mbf{H}_{k}^\mathtt{pri}\big], \label{eqn:post_variance}
\end{align}
where $\mbf{H}_{k}^\mathtt{post} \triangleq \left[\mbf{h}_{k1}^\mathtt{post}, \dots, \mbf{h}_{kM}^\mathtt{post}\right]$.
To summarize, the belief calculation amounts to combining the Gaussian pseudo-likelihood $p(\mbf{H}_{k}^\mathtt{pri}|\mbf{H}_k)$ with the true prior $p(\mbf{H}_k).$
This operation is mathematically equivalent to MMSE denoising under AWGN observations: the pseudo-observation from Module A provides a noisy estimate of the channel, and the prior serves as the denoiser to refine this estimate.
The posterior mean and variance in \eqref{eqn:post_mean} and \eqref{eqn:post_variance} are, however, by no means easy to calculate due to the complicated and high-dimensional integrals.
We will detail the method for obtaining the posterior message in Sections \ref{sec:loss_functions} and \ref{sec:stmp}.

The extrinsic message from the variable node $\mbf{H}_k$ to the factor node $p(\mbf{X}_k|\mbf{H}_k, \alpha_k)$ is calculated as
\begin{align}
	\mathcal{M}_{\mbf{H}_k \rightarrow f_{\mbf{X}_k}}(\mbf{H}_k) &\propto \frac{\mathcal{M}_{f_{\mbf{H}_k}}(\mbf{H}_k) }{\mathcal{M}_{f_{\mbf{X}_k} \rightarrow \mbf{H}_k}(\mbf{H}_k)} \\
	& \propto \prod_{m} \mathcal{CN}(\mbf{h}_{km}; \mbf{h}_{km}^\mathtt{ext}, \tau_{m}^\mathtt{ext}\mbf{I}_N),
\end{align}
where the extrinsic variance and mean are respectively given by
\begin{align}
	\tau_{m}^{\mathtt{ext}} &= \left(\frac{1}{\tau_{m}^{\mathtt{post}}} - \frac{1}{\tau_{m}^{\mathtt{pri}}}\right)^{-1}, \label{eqn:tau_ext}\\
	\mathbf{h}_{km}^\mathtt{ext} &= \tau_{m}^{\mathtt{ext}} \left(\frac{\mathbf{h}_{km}^\mathtt{post}}{\tau_{m}^{\mathtt{post}}} - \frac{\mathbf{h}_{km}^\mathtt{pri}}{\tau_{m}^{\mathtt{pri}}}\right). \label{eqn:H_ext}
\end{align}

\subsubsection{Denoiser of $\alpha_k$}
The belief of $\alpha_k$ on the factor node $p(\mbf{X}_k|\mbf{H}_k, \alpha_k)$ is expressed as
\begin{align}
	&\mathcal{M}_{f_{\mbf{X}_k}}(\alpha_k) \nonumber \\
	&\propto \int_{\mbf{H}_k, \mbf{X}_k} \!\!\!\! \delta (\mbf{X}_k \!-\!\alpha_k \mbf{H}_k) \mathcal{M}_{\mbf{X}_k \rightarrow f_{\mbf{X}_k}} \!\! (\mbf{X}_k) \mathcal{M}_{\mbf{H}_k \rightarrow f_{\mbf{X}_k}} \!\!(\mbf{H}_k) p(\alpha_k) \label{eqn:belief_alpha_dfn}\\
	&\propto (1-\lambda_k^\mathtt{post}) \delta (\alpha_k) + \lambda_k^\mathtt{post} \delta (1-\alpha_k), \label{eqn:belief_alpha}
\end{align}
where
\begin{align}
	&\lambda_k^\mathtt{post} \nonumber \\
	=& \left(1 + \frac{(1-\lambda) \prod_{m} \mathcal{CN} \big(\mbf{0}; \mbf{x}_{km, \mathsf{B}}^\mathtt{pri}, v_{m, \mathsf{B}}^\mathtt{pri} \mbf{I}_N\big)}{\lambda \prod_m \mathcal{CN} \left(\mbf{0}; \mbf{x}_{km, \mathsf{B}}^\mathtt{pri}-\mbf{h}_{km}^\mathtt{ext},  \big(v_{m, \mathsf{B}}^\mathtt{pri} + \tau_{m}^\mathtt{ext} \big)\mbf{I}_N\right)} \right)^{-1} \label{eqn:lambda_post}
\end{align}
is the posterior probability of activity for the $k$-th device.
We set a threshold $\lambda^\mathtt{thr}$ to facilitate the binary decision of device activity, i.e., 
\begin{align}
	\hat{\alpha}_k = \begin{cases}
		1, & \lambda_k^\mathtt{post} \geq \lambda^\mathtt{thr}, \\
		0, & \lambda_k^\mathtt{post} < \lambda^\mathtt{thr}.
	\end{cases} \label{eqn:activity_decision}
\end{align}

\subsubsection{Denoiser of $\mbf{X}_k$}
Based on the update rules of EP \cite{EP2001Minka}, the belief of $\mbf{X}_k$ on the factor node $p(\mbf{X}_k|\mbf{H}_k, \alpha_k)$ is expressed as
\begin{align}
	&\mathcal{M}_{f_{\mbf{X}_k}}(\mbf{X}_k) \nonumber \\
	%%%%%%%%%%%%%%%%%%%%%%%%%%%%%%%%%%%%%%%%%%%%%%%%%%%%%%%%%%%
	&\propto \mathrm{proj} \bigg[  \int_{\mbf{H}_k, \alpha_k} \delta (\mbf{X}_k -\alpha_k \mbf{H}_k) \mathcal{M}_{\mbf{X}_k \rightarrow f_{\mbf{X}_k}}  (\mbf{X}_k) \nonumber \\
	& ~~~~~~~~~~~~~~~~~~~\times  \mathcal{M}_{\mbf{H}_k \rightarrow f_{\mbf{X}_k}} (\mbf{H}_k) p(\alpha_k) \bigg]\nonumber \\
	%%%%%%%%%%%%%%%%%%%%%%%%%%%%%%%%%%%%%%%%%%%%%%%%%%%%%%%%%%%
	&\propto \mathrm{proj} \Big[ (1-\lambda_k^{\mathtt{post}})  \delta (\mbf{X}_k) \nonumber \\
	&~~~~~~ ~~~~+ \lambda_k^{\mathtt{post}} \prod_{m}\mathcal{CN} \big(\mbf{x}_{km}; \tilde{\mbf{x}}_{km, \mathsf{B}}^\mathtt{post}, \tilde{v}_{m,\mathsf{B}}^\mathtt{post} \mbf{I}_N\big) \Big], \label{eqn:X_belief_B}
\end{align}
where
\begin{align}
	\tilde{v}_{m,\mathsf{B}}^{\mathtt{post}} &= \left(\frac{1}{v_{m,\mathsf{B}}^\mathtt{pri}} + \frac{1}{\tau_m^\mathtt{ext}} \right)^{-1}, \\
	\tilde{\mathbf{x}}_{km,\mathsf{B}}^\mathtt{post} &= \tilde{v}_{m,\mathsf{B}}^{\mathtt{post}} 
	\left(\frac{\mathbf{x}_{km,\mathsf{B}}^\mathtt{pri}}{v_{m,\mathsf{B}}^{\mathtt{pri}}} + \frac{\mathbf{h}_{km}^\mathtt{ext}}{\tau_{m}^{\mathtt{ext}}}\right).
\end{align}
In \eqref{eqn:X_belief_B}, the expression inside the Gaussian projection operator is a Bernoulli-Gaussian pdf.
By taking the Gaussian projection, the mean of $\mbf{x}_{km}$ with respect to the belief $\mathcal{M}_{f_{\mbf{X}_k}}(\mbf{X}_k)$ is given by
\begin{align}
		\mathbf{x}_{km,\mathsf{B}}^\mathtt{post} = \lambda_k^\mathtt{post}\tilde{\mathbf{x}}_{km,\mathsf{B}}^\mathtt{post}. \label{eqn:X_B_post}
\end{align}
The variance of the $(n,m)$-th element of $\mbf{X}_k$ with respect to the belief $\mathcal{M}_{f_{\mbf{X}_k}}(\mbf{X}_k)$ is
\begin{align}
	v_{knm,\mathsf{B}}^{\mathtt{post}} = \lambda_k^\mathtt{post} \left(|\tilde{x}_{knm, \mathsf{B}}^\mathtt{post}|^2+ \tilde{v}_{m,\mathsf{B}}^{\mathtt{post}} \right) - |x_{knm, \mathsf{B}}^\mathtt{post}|^2.
\end{align}
Based on the above, the projected Gaussian belief is expressed as
\begin{align}
	\mathcal{M}_{f_{\mbf{X}_k}}(\mbf{X}_k) \approx \prod_{m}\mathcal{CN} \big(\mbf{x}_{km}; \mbf{x}_{km, \mathsf{B}}^\mathtt{post}, v_{m,\mathsf{B}}^\mathtt{post} \mbf{I}_N\big),
\end{align}
where we approximate the variance of each column of $\mbf{X}_k$ by
\begin{align}
	v_{m,\mathsf{B}}^\mathtt{post} = \frac{1}{KN} \sum_{k,n} v_{knm,\mathsf{B}}^{\mathtt{post}}. \label{eqn:v_B_post}
\end{align}
While per-element or full covariance estimation could, in principle, capture heterogeneous confidence across users and subcarriers, such estimates are highly unreliable with limited pilot observations.
Scalar averaging, as adopted in~\eqref{eqn:v_B_post}, is a standard design choice to stabilize message-passing updates \cite{ma2015turbo_cs, vamp2019tit}.

Finally, we calculate the message from the factor node $p(\mbf{X}_k|\mbf{H}_k, \alpha_k)$ to the variable node $\mbf{X}_k$ as
\begin{align}
	\mathcal{M}_{f_{\mbf{X}_k} \rightarrow \mbf{X}_k}(\mbf{X}_k) &= \frac{\mathcal{M}_{f_{\mbf{X}_k}}(\mbf{X}_k)}{\mathcal{M}_{\mbf{X}_k \rightarrow f_{\mbf{X}_k}} (\mbf{X}_k)}, \\
	&= \prod_m \mathcal{CN}\big(\mbf{x}_{km}; \mathbf{x}_{km,\mathsf{B}}^\mathtt{ext}, v_{m,\mathsf{B}}^{\mathtt{ext}}\big), \label{eqn:ext_X_B}
\end{align}
where 
\begin{align}
	v_{m,\mathsf{B}}^{\mathtt{ext}} &= \left(\frac{1}{v_{m,\mathsf{B}}^{\mathtt{post}}} - \frac{1}{v_{m,\mathsf{B}}^{\mathtt{pri}}}\right)^{-1}, \label{eqn:v_B_ext} \\
	\mathbf{x}_{km,\mathsf{B}}^\mathtt{ext} &= v_{m,\mathsf{B}}^{\mathtt{ext}} \left(\frac{\mathbf{x}_{km,\mathsf{B}}^\mathtt{post}}{v_{m,\mathsf{B}}^{\mathtt{post}}} - \frac{\mathbf{x}_{km,\mathsf{B}}^\mathtt{pri}}{v_{m,\mathsf{B}}^{\mathtt{pri}}}\right). \label{eqn:X_B_ext}
\end{align}
The extrinsic message in \eqref{eqn:ext_X_B} flows leftward through the vector-matrix reorganization module unchanged, serving as the prior message of module A in the next iteration, i.e., $\mathbf{x}_{km,\mathsf{A}}^\mathtt{pri} = \mathbf{x}_{km,\mathsf{B}}^\mathtt{ext}$, $v_{m,\mathsf{A}}^{\mathtt{pri}} = v_{m,\mathsf{B}}^{\mathtt{ext}}$.
We summarize the pseudo-code of the TMP framework in Algorithm \ref{alg:framework}.

\begin{algorithm}[t]
	\caption{TMP Framework}
	\label{alg:framework}
	\begin{algorithmic}[1]
		\STATE {\bfseries Input:} $\mbf{Q}$, $\mbf{Y}$, $\delta_0^2$, $\{\mbf{x}_{m, \mathsf{A}}^\mathtt{pri}\}_{m=1}^M$, $\{v_{m, {\sf A}}^\mathtt{pri}\}_{m=1}^M$
		\STATE {\bfseries Output:} $\{\hat{\alpha}_k\}_{k=1}^K$, $\{\mbf{H}_{k}^\mathtt{post}\}_{k=1}^K$
		\REPEAT
		\STATE \% Linear estimator
		\STATE Update $\mbf{x}_{m, \mathsf{A}}^\mathtt{post}$, $v_{m, \mathsf{A}}^\mathtt{post}$ by \eqref{eqn:X_post_A} and \eqref{eqn:v_post_A}, $\forall m$;
		\STATE Update $\mbf{x}_{m, \mathsf{A}}^\mathtt{ext}$, $v_{m, \mathsf{A}}^\mathtt{ext}$ by \eqref{eqn:X_ext_A} and \eqref{eqn:v_ext_A}, $\forall m$;
		\STATE $\mathbf{x}_{km,\mathsf{B}}^\mathtt{pri} = \mathbf{x}_{km,\mathsf{A}}^\mathtt{ext}$, $v_{m,\mathsf{B}}^{\mathtt{pri}} = v_{m,\mathsf{A}}^{\mathtt{ext}}$, $\forall k, m$; \label{alg1:module_B_input}
		\STATE \% Channel and device activity denoisers
		\STATE $\mathbf{h}_{km}^\mathtt{pri} = \mathbf{x}_{km,\mathsf{B}}^\mathtt{pri}$, $\tau_{m}^{\mathtt{pri}} = v_{m,\mathsf{B}}^{\mathtt{pri}}$, $\forall k, m$;
		\STATE Update $\mbf{H}_{k}^\mathtt{post}$, $\tau_{m}^\mathtt{post}$ by \eqref{eqn:post_mean} and \eqref{eqn:post_variance}, $\forall k, m$;
		\STATE Update $\mathbf{h}_{km}^\mathtt{ext}$, $\tau_{m}^\mathtt{ext}$ by \eqref{eqn:H_ext} and \eqref{eqn:tau_ext}, $\forall k, m$;
		\STATE Update $\lambda_k^\mathtt{post}$ by \eqref{eqn:lambda_post}, $\forall k$;
		\STATE Update $\mbf{x}_{km,\mathsf{B}}^\mathtt{post}$, $v_{m, \mathsf{B}}^\mathtt{post}$ by \eqref{eqn:X_B_post} and \eqref{eqn:v_B_post}, $\forall k,m$;
		\STATE Update $\mbf{x}_{km, \mathsf{B}}^\mathtt{ext}$, $v_{m, \mathsf{B}}^\mathtt{ext}$ by \eqref{eqn:X_B_ext} and \eqref{eqn:v_B_ext}, $\forall k,m$;
		\STATE $\mathbf{x}_{km,\mathsf{A}}^\mathtt{pri} = \mathbf{x}_{km,\mathsf{B}}^\mathtt{ext}$, $v_{m,\mathsf{A}}^{\mathtt{pri}} = v_{m,\mathsf{B}}^{\mathtt{ext}}$, $\forall k,m$; \label{alg1:module_A_input}
		\UNTIL{the stopping criterion is met}
		\STATE Decide the indices of active devices by \eqref{eqn:activity_decision}, $\forall k$.
	\end{algorithmic}
\end{algorithm}

\section{Learning Score-Based Generative Priors for MMSE Denoising} \label{sec:loss_functions}
In this section, we introduce score-based representations of \eqref{eqn:post_mean} and \eqref{eqn:post_variance} to circumvent the intractable high-dimensional integrals involved in posterior message calculation.
Then, we elaborate the denoising score matching technique \cite{vincent2011connection, song2019generative, lu2022maximum}, a data-driven approach for empirically estimating score functions.
\subsection{Bridging Score Function and MMSE Denoising} \label{sec:bridging}
For simplicity, we approximate that in \eqref{eqn:awgn_observation}, different columns of $\mbf{H}_k^\mathtt{pri}$ (i.e., $\mbf{h}_{km}^\mathtt{pri}$) share a common variance given by $\tau^\mathtt{pri} = \frac{1}{M} \sum_{m} \tau_m^\mathtt{pri}$.
Then, we rewrite the AWGN observation model as
\begin{align} \label{eqn:awgn_matrix}
	\mbf{H}_k^\mathtt{pri} =  \mbf{H}_{k} + \mbf{W}_{k}, ~~ \mbf{W}_{k} \sim \mathcal{CN}\big(\mbf{0}, \tau^\mathtt{pri} \mbf{I}_{NM}\big) .
\end{align}
For the model in \eqref{eqn:awgn_matrix}, Tweedie's formula \cite{robbins1992empirical, efron2011tweedie} offers an empirical Bayes approach to calculate the posterior mean, applicable regardless of the specific form of the prior distribution $p(\mbf{H}_k)$.
The formula reads
\begin{align}\label{eqn:tweedie}
	\mathbb{E}\big[\mbf{H}_k|\mbf{H}_{k}^\mathtt{pri}\big] = \mbf{H}_{k}^\mathtt{pri} + \tau^\mathtt{pri} \nabla_{\mbf{H}_{k}^\mathtt{pri}} \log p(\mbf{H}_{k}^\mathtt{pri}),
\end{align}
where $\nabla_{\mbf{H}_{k}^\mathtt{pri}} \log p(\mbf{H}_{k}^\mathtt{pri}) \in \mathbb{C}^{N \times M}$ is the gradient of the log-density function, commonly known as the (first-order) score function.
Tweedie's formula bridges MMSE denoising and the score of the perturbed data distribution through a simple equation, eliminating the need for calculating high-dimensional integrals.
Specifically, the posterior mean can be viewed as the sum of the maximum likelihood estimator and a Bayes correction term.
The formula can also be framed as ``denoised result = noisy data - predicted noise'' from a denoising perspective, in which the predicted noise corresponds to $-\tau^\mathtt{pri} \nabla_{\mbf{H}_{k}^\mathtt{pri}} \log p(\mbf{H}_{k}^\mathtt{pri})$.

Likewise, the posterior covariance matrix can be obtained through the second-order Tweedie's formula \cite{efron2011tweedie} as
\begin{align}\label{eqn:tweedie_2nd}
	\!\!\mathrm{Cov}\big[\mbf{H}_k|\mbf{H}_{k}^\mathtt{pri}\big] \!=\! \tau^\mathtt{pri} \mbf{I}_{NM} \!+ (\tau^\mathtt{pri})^2 \nabla^2_{\mbf{H}_{k}^\mathtt{pri}} \log p(\mbf{H}_{k}^\mathtt{pri}),
\end{align}
where $\nabla_{\mbf{H}_{k}^\mathtt{pri}}^2 \log p(\mbf{H}_{k}^\mathtt{pri}) \in \mathbb{C}^{NM \times NM}$ is the Hessian of the log-density function, referred to as the second-order score function.
The first- and second-order Tweedie's formulas are well-known results that can be proven using simple integration tricks \cite[Appendix B]{tweedies_proof2024nips} or by applying the properties of exponential families \cite[Appendix A]{finzi2023user}.

Once the score functions are known, we can apply \eqref{eqn:tweedie} and \eqref{eqn:tweedie_2nd} to efficiently implement the MMSE denoiser described in \eqref{eqn:post_mean} and \eqref{eqn:post_variance}.
For massive connectivity, learning the score functions individually for each device is impractical due to the vast number of devices involved.
Instead, we train only two networks, namely the first- and the second-order score networks, in which all the training samples from different devices are combined into a single training set.

\subsection{First-Order Score Matching}
With a slight abuse of notation, let $p(\mbf{H}) : \mathbb{C}^{N \times M}\rightarrow \mathbb{R}$ be the mixture distribution  of the channels for the total $K$ devices.
We perturb it with a pre-specified noise distribution $p(\tilde{\mbf{H}}|\mbf{H}) = \mathcal{CN}(\mbf{0}, \tau \mbf{I}_{NM})$.
Consequently, the perturbed channel distribution is expressed as $p(\tilde{\mbf{H}}) = \int_{\mbf{H}}  p(\tilde{\mbf{H}}|\mbf{H}) p(\mbf{H})$.
For a specific value of $\tau$, our goal is to learn a first-order score network $\mbf{S}_1(\cdot, \tau; \bsm{\theta}): \mathbb{C}^{N\times M} \rightarrow \mathbb{C}^{N\times M}$ parameterized by $\bsm{\theta}$ with the following score matching objective:
\begin{align} \label{eqn:1st_score_matching}
	\min_{\bsm{\theta}} ~\mathbb{E}_{p(\tilde{\mbf{H}})}\left[ \left\|\mbf{S}_{1} (\tilde{\mbf{H}}, \tau; \bsm{\theta}) - \nabla_{\tilde{\mbf{H}}} \log p(\tilde{\mbf{H}}) \right\|^2_F \right].
\end{align}
Eqn. \eqref{eqn:1st_score_matching} cannot be directly used as the loss function since $\nabla_{\tilde{\mbf{H}}} \log p(\tilde{\mbf{H}})$ does not have an analytical form.
The above formulation is shown equivalent to the denoising score matching objective \cite{vincent2011connection, song2019generative}, given by
\begin{align} \label{eqn:1st_dsm}
	\min_{\bsm{\theta}}~\mathbb{E}_{p(\mbf{H})p(\tilde{\mbf{H}}|\mbf{H})} \left[ \left\|\mbf{S}_{1} (\tilde{\mbf{H}}, \tau; \bsm{\theta}) - \nabla_{\tilde{\mbf{H}}} \log p(\tilde{\mbf{H}}|\mbf{H}) \right\|^2_F \right].
\end{align}
The equivalence is in the sense that \eqref{eqn:1st_score_matching} and \eqref{eqn:1st_dsm} share the same optimal solution $\mbf{S}_{1} (\tilde{\mbf{H}}, \tau; \bsm{\theta}^\star) = \nabla_{\tilde{\mbf{H}}} \log p(\tilde{\mbf{H}})$, supposing that the network has sufficient representation capability.
Substituting $\nabla_{\tilde{\mbf{H}}} \log p(\tilde{\mbf{H}}|\mbf{H}) = -\frac{\tilde{\mbf{H}} - \mbf{H}}{\tau}$ into \eqref{eqn:1st_dsm} yields
\begin{align} \label{1st_dsm_2}
	\min_{\bsm{\theta}}~ \ell_1 (\tau; \bsm{\theta}) \triangleq  \mathbb{E}_{p(\mbf{H})p(\tilde{\mbf{H}}|\mbf{H})} \left[ \left\|\mbf{S}_{1} (\tilde{\mbf{H}}, \tau; \bsm{\theta}) + \frac{\tilde{\mbf{H}} - \mbf{H}}{\tau} \right\|^2_F \right].
\end{align}
Eqn. \eqref{1st_dsm_2} implies that the first-order score matching is essentially to learn a noise predictor that outputs the noise scaled by a negative factor.

In practice, the score-based MMSE denoiser is required to operate effectively across a range of noise levels.
We therefore combine \eqref{1st_dsm_2} for all possible values of $\tau \in \{\tau_i\}_{i=1}^I$ to establish a unified objective for training the first-order score network:
\begin{align} \label{eqn:1st_loss}
	\min_{\bsm{\theta}} ~ \sum_{i=1}^{I} \lambda_1(\tau_i)  \ell_1 (\tau_i; \bsm{\theta}).
\end{align}
A common choice for the weighting factor is $\lambda_1(\tau_i) = \tau_i$~\cite{song2019generative}, which empirically balances the loss term $\ell_1 (\tau_i; \bsm{\theta})$ for different values of $\tau_i$.

\subsection{Second-Order Score Matching}
For a given $\tau$, we seek to learn a second-order score model $\mbf{S}_{2}(\cdot, \tau; \bsm{\phi}): \mathbb{C}^{N\times M} \rightarrow \mathbb{C}^{NM \times NM}$ parameterized by $\bsm{\phi}$ with the second-order score matching objective:
\begin{align} \label{2nd_sm}
	\min_{\bsm{\phi}}~\mathbb{E}_{p(\tilde{\mbf{H}})} \left[ \left\|\mbf{S}_{2}(\tilde{\mbf{H}}, \tau; \bsm{\phi}) - \nabla_{\tilde{\mbf{H}}}^2 \log p(\tilde{\mbf{H}}) \right\|^2_F \right].
\end{align}
The second-order score matching objective can also be expressed in an equivalent denoising score matching formulation as
\begin{align} 
	\min_{\bsm{\phi}}~&\mathbb{E}_{p(\mbf{H})p(\tilde{\mbf{H}}|\mbf{H})} \bigg[ \Big\|\mbf{S}_{2}(\tilde{\mbf{H}}, \tau; \bsm{\phi}) \nonumber \\
	&~~~~~~~~~- \mbf{b}(\mbf{H}, \tilde{\mbf{H}}, \tau) \mbf{b}(\mbf{H},\tilde{\mbf{H}}, \tau)^{\sf H} +  \frac{\mbf{I}_{NM}}{\tau}\Big\|_F^2\bigg], \label{2nd_dsm}
\end{align}
where $\mbf{b}(\mbf{H}, \tilde{\mbf{H}}, \tau) \triangleq \mathrm{vec} \left(\nabla_{\tilde{\mbf{H}}} \log p(\tilde{\mbf{H}}) + \frac{\tilde{\mbf{H}} - \mbf{H}}{\tau} \right)$.
Please refer to \cite[Appendix F]{lu2022maximum} for the detailed derivation.
The objective in \eqref{2nd_dsm} relies on the ground truth of the first-order score function, which is in general unavailable.
To address this, we replace $\nabla_{\tilde{\mbf{H}}} \log p(\tilde{\mbf{H}})$ with the learned model $\mbf{S}_{1}(\tilde{\mbf{H}}, \tau; \bsm{\theta})$ for efficient computation.
Consequently, the objective in \eqref{2nd_dsm} modifies to
\begin{align} 
	\min_{\bsm{\phi}}~&\mathbb{E}_{p(\mbf{H})p(\tilde{\mbf{H}}|\mbf{H})} \bigg[ \Big\|\mbf{S}_{2}(\tilde{\mbf{H}}, \tau; \bsm{\phi}) \nonumber \\
	&~~~~~~~~~- \hat{\mbf{b}}(\mbf{H}, \tilde{\mbf{H}}, \tau) \hat{\mbf{b}}(\mbf{H}, \tilde{\mbf{H}}, \tau)^{\sf H} +  \frac{\mbf{I}_{NM}}{\tau}\Big\|_F^2\bigg], \label{2nd_dsm_estimated}
\end{align}
where $\hat{\mbf{b}}(\mbf{H}, \tilde{\mbf{H}}, \tau) \triangleq \mathrm{vec} \left(\mbf{S}_{1}(\tilde{\mbf{H}}, \tau; \bsm{\theta}) + \frac{\tilde{\mbf{H}} - \mbf{H}}{\tau} \right)$.
It has been established in \cite{lu2022maximum} that this replacement allows the second-order score network to maintain an error-bounded property, provided that the first-order score matching error remains bounded.

As demonstrated in \eqref{eqn:post_variance}, only the trace of the posterior covariance matrix is required, rather than the full matrix itself.
By noting in \eqref{eqn:tweedie_2nd} the simple relation between the posterior covariance matrix and the second-order score function, 
it suffices to match only the trace of the second-order score function for posterior variance evaluation.
This leads to a simplified objective expressed as
\begin{align} \label{2nd_dsm_trace}
	\min_{\bsm{\phi}}~ \ell_2(\tau; \bsm{\phi}) &\triangleq \mathbb{E}_{p(\mbf{H})p(\tilde{\mbf{H}}|\mbf{H})} \bigg[ \bigg|\mathrm{tr}\big(\mbf{S}_{2}(\tilde{\mbf{H}}, \tau; \bsm{\phi})\big) \nonumber \\
	&~~~~~~~~~~~- \left\|\hat{\mbf{b}}(\mbf{H}, \tilde{\mbf{H}}, \tau)\right\|_2^2 +  \frac{NM}{\tau}\bigg|^2\bigg].
\end{align}
We combine \eqref{2nd_dsm_trace} for all $\tau \in \{\tau_i\}_{i=1}^I$ to formulate a unified objective for training the second-order score network:
\begin{align} \label{2nd_dsm_unified}
	\min_{\bsm{\phi}}~ \sum_{i=1}^{I} \lambda_2(\tau_i)  \ell_2(\tau_i; \bsm{\phi}),
\end{align}
where $\lambda_2(\tau_i)$ is the weighting factor depending on $\tau_i$.
Following~\cite{lu2022maximum}, we choose $\lambda_2(\tau_i) = \tau_i^2$.

In practice, one can first train a first-order score network, and then freeze it during the second-order score learning phase.
Alternatively, both networks can be trained simultaneously by disabling gradient backpropagation from the second-order training objective \eqref{2nd_dsm_unified} to the first-order score network.

\begin{figure*}
	[t]
	\centering
%		\vspace{-.5em}
	\includegraphics[width=2\columnwidth]{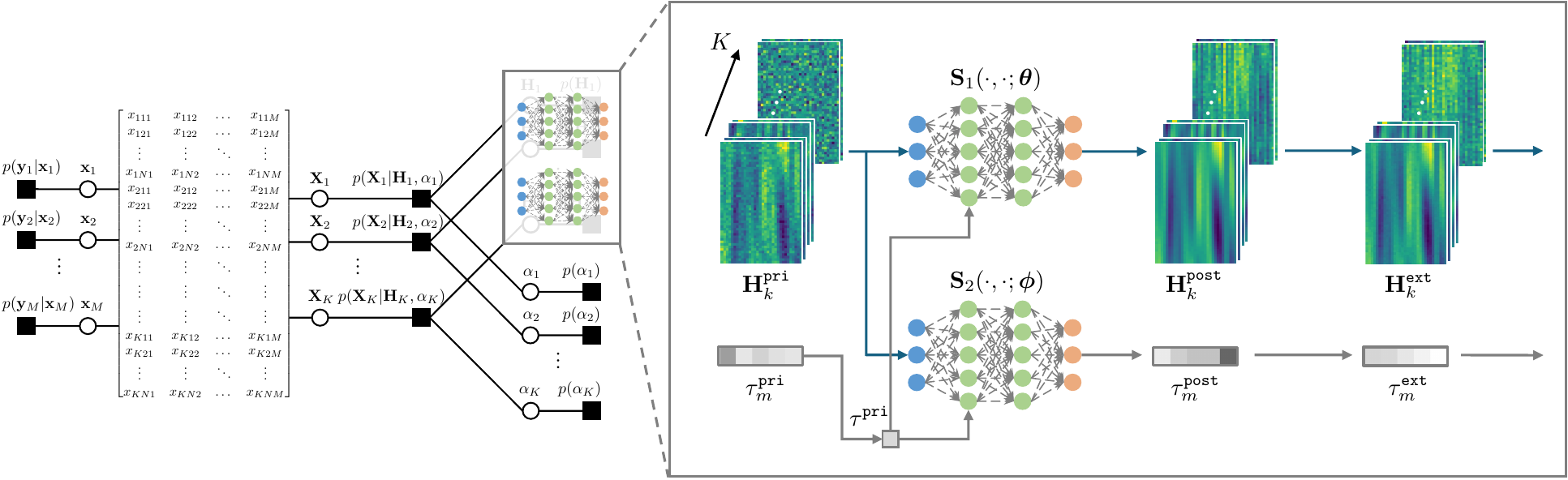}
	%	\vspace{-.5em}
	\caption{Diagram of the STMP-JADCE algorithm.
	The real and imaginary parts of the wireless channel matrix are concatenated as two feature channels (analogous to the RGB channels in images) when fed into the score networks.
    For the $k$-the device, the input tensor is constructed by stacking the real and imaginary parts of $\mbf{H}_k$ into a tensor of size $(N,M,2)$.
    Across multiple devices, these tensors are further batched along the user dimension.
	In this figure, the first and the second devices are active, whereas the $K$-th device is inactive.
	Therefore, the input to the score networks takes the form of $\mbf{H}_k^\mathtt{pri} = \mbf{H}_{k} + \mbf{W}_{k}$ for $k=1,2$, while $\mbf{H}_K^\mathtt{pri}$ is a random noise.
	Note that, for brevity, we omitted the skip connections (dependences) from the prior messages to both the the posterior and extrinsic messages.
	Please refer to eqns. \eqref{eqn:stmp_post_mean}, \eqref{eqn:stmp_post_var}, \eqref{eqn:tau_ext}, and \eqref{eqn:H_ext} for the rigorous relationships.} 
	\label{fig:factor_graph_score}
%		\vspace{-.5em}
\end{figure*}

\section{STMP-JADCE Algorithm} \label{sec:stmp}
\subsection{Overall Algorithm}
With the two score networks available, we are able to obtain the posterior mean $\mbf{H}_{k}^\mathtt{post}$ and variance $\tau_m^\mathtt{post}$ based on \eqref{eqn:tweedie} and \eqref{eqn:tweedie_2nd} as
\begin{align}
	\mbf{H}_{k}^\mathtt{post} &= \mbf{H}_{k}^\mathtt{pri} + \tau^\mathtt{pri} \mbf{S}_{1} \big(\mbf{H}_{k}^\mathtt{pri}, \tau^\mathtt{pri}; \bsm{\theta}\big), \label{eqn:stmp_post_mean}\\
	\tau_m^\mathtt{post} &= \tau^\mathtt{pri} + \frac{(\tau^\mathtt{pri})^2}{KN} \sum_{k,n} s_{2, nm} \big(\mbf{H}_{k}^\mathtt{pri}, \tau^\mathtt{pri}; \bsm{\phi}\big) \label{eqn:stmp_post_var},
\end{align}
where $s_{2, nm} (\mbf{H}_{k}^\mathtt{pri}, \tau^\mathtt{pri}; \bsm{\phi})$ denotes the $((n-1)M +  m)$-th diagonal elements of $\mbf{S}_{2} \big(\mbf{H}_{k}^\mathtt{pri}, \tau^\mathtt{pri}; \bsm{\phi}\big)$.

Fig. \ref{fig:factor_graph_score} demonstrates the integration of the two score networks into the TMP framework, resulting in the STMP-JADCE algorithm.
STMP-JADCE features a hybrid structure in which the channel denoiser is realized by score networks, while the device activity denoiser and the linear estimator are derived analytically.
In this way, our method can be easily applied across different device activity levels, pilot patterns, and SNRs without the need to retain the score networks.
Moreover, the potential mismatch between the chosen prior of the device activity and the ground truth can be rectified by incorporating the expectation-maximization (EM) update \cite{BiGAMP2014tsp}.
Essentially, the development of STMP-JADCE highlights a broader philosophical shift: whereas conventional Bayesian inference in wireless communications has long relied on manually specified and typically oversimplified priors for tractability, our framework embraces a data-driven paradigm in which the channel prior is learned directly from data. This data-driven Bayesian inference not only resolves the mismatch between hand-crafted priors and real channel statistics, but also opens the door to more principled integration of modern generative models into inference algorithms.
We discuss the engineering tricks used in STMP-JADCE in the following.

\subsection{Engineering Tricks}
\subsubsection{Damping}
The pseudo AWGN observation model in \eqref{eqn:awgn_observation} is justified only in the large system limit with right-rotationally invariant random $\mbf{Q}$ \cite{se_analysis2018nips}.
In practice, the equivalent noise may not be strictly Gaussian due to the particular choice of $\mbf{Q}$ and the finite dimensions of $M$, $N$, $K$, and $T$.
This discrepancy can lead to some unexpected numerical issues and diverges.
In this work, we employ the damping technique to improve the numerical stability of STMP-JADCE.
Specifically, in each iteration, we smoothen the updates of the means and variances by using a linear combination of the current and previous updates.
For example, we replace the input to module A (Line \ref{alg1:module_A_input} of Algorithm \ref{alg:framework}) at the $i$-th iteration by
\begin{align}
	\mathbf{x}_{km,\mathsf{A}}^\mathtt{pri}(i) &=  \gamma \mathbf{x}_{km,\mathsf{B}}^\mathtt{ext} (i) + (1-\gamma) \mathbf{x}_{km,\mathsf{B}}^\mathtt{ext} (i-1), \\
	v_{m,\mathsf{A}}^{\mathtt{pri}}(i) &=  \gamma v_{m,\mathsf{B}}^{\mathtt{ext}} (i) + (1-\gamma) v_{m,\mathsf{B}}^{\mathtt{ext}} (i-1),
\end{align}
where $\gamma \in (0, 1]$ is the damping factor.
The input to module B is damped in a similar manner. 

\begin{algorithm}[t]
	\caption{STMP-JADCE Algorithm}
	\label{alg:stmp}
	\begin{algorithmic}[1]
		\STATE {\bfseries Input:} $\mbf{Q}$, $\mbf{Y}$, $\delta_0^2$, $\{\mbf{x}_{m, \mathsf{A}}^\mathtt{pri}\}_{m=1}^M$, $\{v_{m, {\sf A}}^\mathtt{pri}\}_{m=1}^M$
		\STATE {\bfseries Output:} $\{\hat{\alpha}_k\}_{k=1}^K$, $\{\mbf{H}_{k}^\mathtt{post}\}_{k=1}^K$
		\REPEAT
		\STATE \% Linear estimator
		\STATE Update $\mbf{x}_{m, \mathsf{A}}^\mathtt{post}(i)$, $v_{m, \mathsf{A}}^\mathtt{post}(i)$ by \eqref{eqn:X_post_A} and \eqref{eqn:v_post_A}, $\forall m$;
		\STATE Update $\mbf{x}_{m, \mathsf{A}}^\mathtt{ext}(i)$, $v_{m, \mathsf{A}}^\mathtt{ext}(i)$ by \eqref{eqn:X_ext_A} and \eqref{eqn:v_ext_A}, $\forall m$;
		\STATE $\mathbf{x}_{km,\mathsf{B}}^\mathtt{pri}(i) = \gamma \mathbf{x}_{km,\mathsf{A}}^\mathtt{ext} (i) + (1-\gamma) \mathbf{x}_{km,\mathsf{A}}^\mathtt{ext} (i-1)$,
		$v_{m,\mathsf{B}}^{\mathtt{pri}}(i) = \gamma v_{m,\mathsf{A}}^{\mathtt{ext}} (i) + (1-\gamma) v_{m,\mathsf{A}}^{\mathtt{ext}} (i-1)$, $\forall k, m$;
		\STATE \% Channel and device activity denoisers
		\STATE $\mathbf{h}_{km}^\mathtt{pri}(i) = \mathbf{x}_{km,\mathsf{B}}^\mathtt{pri}(i)$,
		$\tau_{m}^{\mathtt{pri}}(i)= v_{m,\mathsf{B}}^{\mathtt{pri}}(i)$, $\forall k, m$;
		\STATE Update $\bar{\mbf{H}}_{k}^\mathtt{pri}(i)$, $\bar{\tau}^\mathtt{pri}(i)$ by \eqref{scaled_pri_mean} and \eqref{scaled_pri_variance}, $\forall k$;
		\STATE Update $\bar{\mbf{H}}_{k}^\mathtt{post}(i)$, $\bar{\tau}_m^\mathtt{post}(i)$ by \eqref{scaled_post_mean} and \eqref{scaled_post_variance}, $\forall k, m$;
		\STATE Update $\mbf{H}_{k}^\mathtt{post}\!(i)$ and $\tau_m^\mathtt{post}\!(i)$ by \eqref{rescaled_post_mean} and \eqref{rescaled_post_variance}, $\!\forall k, \!m$;
		\STATE Update $\mathbf{h}_{km}^\mathtt{ext}(i)$, $\tau_{m}^\mathtt{ext}(i)$ by \eqref{eqn:H_ext} and \eqref{eqn:tau_ext}, $\forall k, m$;
		\STATE Update $\lambda_k^\mathtt{post}(i)$ by \eqref{eqn:lambda_post}, $\forall k$;	
		\STATE Update $\mbf{x}_{km,\mathsf{B}}^\mathtt{post}(i)$, $v_{m, \mathsf{B}}^\mathtt{post}(i)$ by \eqref{eqn:X_B_post} and \eqref{eqn:v_B_post}, $\forall k,m$;
		\STATE Update $\mbf{x}_{km, \mathsf{B}}^\mathtt{ext} (i)$, $v_{m, \mathsf{B}}^\mathtt{ext} (i)$ by \eqref{eqn:X_B_ext} and \eqref{eqn:v_B_ext}, $\forall k,m$;
		\STATE $\mathbf{x}_{km,\mathsf{A}}^\mathtt{pri}(i) = \gamma \mathbf{x}_{km,\mathsf{B}}^\mathtt{ext} (i) + (1-\gamma) \mathbf{x}_{km,\mathsf{B}}^\mathtt{ext} (i-1)$,
		$v_{m,\mathsf{A}}^{\mathtt{pri}}(i) = \gamma v_{m,\mathsf{B}}^{\mathtt{ext}} (i) + (1-\gamma) v_{m,\mathsf{B}}^{\mathtt{ext}} (i-1)$, $\forall k, m$;
		\UNTIL{the stopping criterion is met}
		\STATE Decide the indices of active devices by \eqref{eqn:activity_decision}, $\forall k$.
	\end{algorithmic}
\end{algorithm}

\subsubsection{Channel Power Normalization}
Following the training recipe of score-based generative models \cite{song2019generative}, we normalize the power of each channel sample during training, ensuring that the normalized channel $\bar{\mbf{H}}_k = \sqrt{NM} \mbf{H}_k / \left\|\mbf{H}_k\right\|_F$.
During inference, the ground-truth channel power is unknown due to channel fading.
The input should be properly scaled since the networks are trained on a specific channel power and may not generalize well to arbitrary power levels.
According to \eqref{eqn:awgn_matrix}, we have
\begin{align} \label{eqn:power_relation}
	\big\|\mbf{H}_k\big\|_F^2 = \big\|\mbf{H}_k^\mathtt{pri}\big\|_F^2 - NM \tau^\mathtt{pri}.
\end{align}
By comparing \eqref{eqn:power_relation} with the training normalization $\left\|\bar{\mbf{H}}_k\right\|_F^2 = NM$, we compute the scaled input of $\mbf{H}_k^\mathtt{pri}$ as follows:
\begin{align}
	\bar{\mbf{H}}_k^\mathtt{pri} = \sqrt{\frac{NM}{\big\|\mbf{H}_k^\mathtt{pri}\big\|_F^2 - NM\tau^\mathtt{pri}}} \mbf{H}_k^\mathtt{pri}. \label{scaled_pri_mean}
\end{align}
Similarly, the scaled input of $\tau^\mathtt{pri}$ is given by
\begin{align}
	\bar{\tau}^\mathtt{pri} = \frac{KNM}{\sum_{k} \big\|\mbf{H}_k^\mathtt{pri}\big\|_F^2 - KNM\tau^\mathtt{pri}} \tau^\mathtt{pri}. \label{scaled_pri_variance}
\end{align}
Let $\bar{\mbf{H}}_{k}^\mathtt{post}$ and $\bar{\tau}_{m}^\mathtt{post}$ be the scaled posterior mean and variance obtained from the score networks:
\begin{align}
	\bar{\mbf{H}}_{k}^\mathtt{post} &= \bar{\mbf{H}}_{k}^\mathtt{pri} + \bar{\tau}^\mathtt{pri} \mbf{S}_{1} \big(\bar{\mbf{H}}_{k}^\mathtt{pri}, \bar{\tau}^\mathtt{pri}; \bsm{\theta}\big), \label{scaled_post_mean} \\
	\bar{\tau}_{m}^\mathtt{post} &= \bar{\tau}^\mathtt{pri} + \frac{(\bar{\tau}^\mathtt{pri})^2}{KN} \sum_{k,n} s_{2, nm} \big(\bar{\mbf{H}}_{k}^\mathtt{pri}, \bar{\tau}^\mathtt{pri}; \bsm{\phi}\big). \label{scaled_post_variance}
\end{align}
Finally, we rescale $\bar{\mbf{H}}_{k}^\mathtt{post}$ and $\bar{\tau}_{m}^\mathtt{post}$ as
\begin{align}
	\mbf{H}_{k}^\mathtt{post} &= \sqrt{\frac{\big\|\mbf{H}_k^\mathtt{pri}\big\|_F^2 - NM\tau^\mathtt{pri}}{NM}} \bar{\mbf{H}}_{k}^\mathtt{post}, \label{rescaled_post_mean}\\
	\tau_m^\mathtt{post} &=\frac{\sum_{k}\big\|\mbf{H}_k^\mathtt{pri}\big\|_F^2 - KNM\tau^\mathtt{pri}}{KNM} \bar{\tau}_m^\mathtt{post}. \label{rescaled_post_variance}
\end{align}

Algorithm \ref{alg:stmp} provides the pseudo-code of STMP-JADCE with all the engineering tricks implemented.

\subsection{Pilot Design and Complexity Analysis\label{sec:pilot_design}}
Recall that we assume partial orthogonal $\mbf{Q}$ to simplify the LMMSE estimator in module A.
To fulfill this requirement, the pilot symbols $\{q_{ktn}\}_{k=1}^K$ of all devices transmitted on the $n$-th subcarrier during the $t$-th OFDM symbol should satisfy
\begin{align}
	\left[q_{1tn}, q_{2tn}, \dots, q_{Ktn}\right] = \sqrt{KP}\mbf{u}_i,
\end{align}
where $\mbf{u}_i \in \mathbb{C}^{1\times K}$ is a row vector randomly selected from a unitary matrix $\mbf{U} \in \mathbb{C}^{K\times K}$.
Therefore, a total of $NT$ rows should be selected from a $K$-by-$K$ matrix.
For a particular $n$, the selected $T$ rows must be distinct for each $t$.
For a particular $t$, the selected $N$ rows do not need to be unique across different values of $n$.
It can now be readily verified that $\mbf{Q}\mbf{Q}^{\sf H} = KP \mbf{I}_{NT}$.
Moreover, if $\mbf{U}$ is a DFT matrix, the matrix multiplications in module A can be simplified through the fast Fourier transform (FFT).

The computational complexity of STMP-JADCE is evaluated in terms of the number of complex multiplications.
The main contributors are the matrix-vector multiplications in module A and the forward passes through the score networks in module B.
In particular, each matrix-vector multiplication in~\eqref{eqn:X_post_A} can be implemented by $N$ length-$K$ FFT/IFFT operations, resulting in a total of $2MNK\log_2 K$ complex multiplications across all $M$ antennas.
Other element-wise operations are of lower order and therefore neglected. 
The complexity of the score networks is denoted by $C = \max \{C_1, C_2\}$, where $C_1$ and $C_2$ represent the computational costs of a forward pass through the first- and second-order score networks, respectively.
Consequently, the overall per-iteration complexity of STMP-JADCE is $\mathcal{O}\left(MNK\log_2 K + C\right)$.

\section{Simulation Results}\label{sec:simulations}
\subsection{Simulation Settings}
\subsubsection{Channel Dataset}
We adopt the CDL family of channel models generated from the QuaDRiGa toolbox \cite{quadriga} for training and evaluation.
The training and evaluation sets consist of $50,000$ and $5,000$ channel samples, respectively.
The evaluation samples are directly generated from the CDL-C model specified in \cite[Table 7.7.1-3]{3GPP_TR_38_901}.
In the training set, we randomly generate the angles, delays, and complex gains of each path in CDL-C to cover a diverse range of channel realizations.
The CDL-C channel model comprises $24$ clusters, each contributing $20$ sub-paths, resulting in a total of $L=480$ propagation paths.

In simulation, the BS is equipped with a uniform linear array (ULA) consisting of $M=32$ antennas, arranged with half-wavelength spacing.
A total of $N=48$ OFDM subcarriers is considered, operating in a central frequency of $2.6$ GHz.
The subcarrier spacing is $\Delta f = 30$ kHz.
The delay spread is randomly chosen in the range of $[100,363]$ ns.
The channel model can be expressed as $\mbf{H} = \sum_{l=1}^L \beta_l \sqrt{\rho_l} \mbf{a}_{N} (\tau_l) \mbf{b}_{M}^{\sf H} (\varphi_l)$, where $\beta_l$, $\rho_l$, $\tau_l$, and $\varphi_l$ denote the complex coefficient, the path power, the path delay, and the angle of arrival (AoA) of the $l$-th path, respectively;
the temporal and spatial steering vectors are defined as $\mbf{a}_N(\tau_l) \triangleq \frac{1}{\sqrt{N}}\left[1, e^{-j2\pi \Delta f \tau_l}, \dots, e^{-j2\pi(N-1) \Delta f \tau_l}\right]^{\sf T}$ and 
$\mbf{b}_M(\varphi_l) \triangleq \frac{1}{\sqrt{M}}\left[1, e^{-j\pi \sin \varphi_l}, \dots, e^{-j\pi (M-1) \sin \varphi_l}\right]^{\sf T}$.
The distance between the devices and the BS ranges from $35$ to $200$ m.
The path-loss component is modeled as $-128.1 - 36.7 \log_{10} d_k$ in dB scale, where $d_k$ is the distance (in km) between the $k$-th device and the BS.

\subsubsection{Network Architectures and Hyperparameters}
We adopt the NCSNv2 \cite{improved_tech2020nips} and NCSN \cite{song2019generative} architectures for the first- and second-order score networks, respectively.
Both networks adopt multi-scale convolutional architectures based on a U-Net or RefineNet backbone, composed of stacked residual and refinement blocks.
Each block consists of two $3\times3$ convolutions followed by instance normalization and nonlinear activation, with skip connections between corresponding encoder and decoder stages to preserve fine-grained spatial information. 
The networks progressively downsample feature maps from $128$ to $256$ channels through residual blocks, then upsample them symmetrically using refinement blocks with dilated convolutions to enlarge the receptive field. 
The noise level is embedded through a conditioning mechanism implemented by a small multilayer perceptron followed by feature-wise linear modulation, which injects noise-dependent scale and shift parameters into every block.
The output of the NCSN model corresponds to the diagonal entries of the second-order score (Hessian) function, sharing the same dimension as the first-order score.
The detailed layer configurations of both architectures are summarized in Table~\ref{tab:net_arch}.

The models are trained on every integer SNR in the range of $[-10, 30]$ dB, 
yielding $I=41$ distinct noise levels $\{\tau_i\}_{i=1}^{I}$.
Both networks are optimized using Adam with a learning rate of $10^{-4}$, batch size $400$, and $400$ training epochs.
An exponential moving average (EMA) of the model parameters with rate $0.999$ is applied to stabilize training~\cite{improved_tech2020nips}.
The training is conducted on a Linux server with 4 NVIDIA A800 GPUs and an Intel Xeon Platinum 8358P CPU @ 2.60 GHz.
Under this configuration, the training of the first- and second-order score networks each takes approximately one day.

\begin{table}[t]
\centering
\caption{Network Architectures}
\label{tab:net_arch}
\begin{tabular}{cc}
\toprule
\makecell{First-order score network \\ (NCSNv2)} & \makecell{Second-order score network \\ (NCSN)} \\
\midrule
3$\times$3 Conv2D, 128 & 3$\times$3 Conv2D, 128 \\
ResBlock, 128 & CondResBlock, 128 \\
ResBlock, 128 & CondResBlock, 128 \\
ResBlock down, 256 & CondResBlock down, 256 \\
ResBlock, 256 & CondResBlock, 256 \\
ResBlock down, 256 (dilation 2) & CondResBlock down, 256 (dilation 2) \\
ResBlock, 256 (dilation 2) & CondResBlock, 256 (dilation 2) \\
ResBlock down, 256 (dilation 4) & CondResBlock down, 256 (dilation 4) \\
ResBlock, 256 (dilation 4) & CondResBlock, 256 (dilation 4) \\
RefineBlock, 256 & CondRefineBlock, 256 \\
RefineBlock, 256 & CondRefineBlock, 256 \\
RefineBlock, 128 & CondRefineBlock, 128 \\
RefineBlock, 128 & CondRefineBlock, 128 \\
3$\times$3 Conv2D, 2 & 3$\times$3 Conv2D, 2 \\
\bottomrule
\end{tabular}
\end{table}

\subsubsection{Evaluation Settings}
During evaluation, the channel of each device is randomly sampled from the evaluation set.
The following settings are adopted unless specified otherwise: $\lambda = 0.05$, $K= 800$, and the SNR is $10$ dB.
For the figures showing the NMSE\footnote{The NMSE of channel estimation is defined as $\frac{\sum_k\| \mbf{H}_k - \mbf{H}_k^\mathtt{post}\|_F^2 }{\sum_k\| \mbf{H}_k\|_F^2 }$.} performance and detection error probability, each data point is averaged over $500$ and $5000$ Monte Carlo trials, respectively.
The evaluation utilizes a single NVIDIA A800 GPU, which is performed on the same machine used for training.

\subsection{Score-Based MMSE Denoising Performance}
Before digging into the evaluation of the STMP-JADCE algorithm, we separately assess the performance of the score-based MMSE denoiser specified in \eqref{eqn:tweedie}.
We consider the task of recovering $\mbf{H} \in \mathbb{C}^{N\times M}$ from its AWGN-corrupted observation $\mbf{Y} = \mbf{H} + \mbf{N}$.
We assume for simplicity that the path-loss are perfectly compensated.
The following baselines are considered:
\begin{itemize}
	\item \textbf{Least square (LS):}
	The LS estimator is $\hat{\mbf{H}}_{\mathrm{LS}} = \mbf{Y}$.
	\item \textbf{Orthogonal matching pursuit (OMP) \cite{omp2011tit}:} A greedy algorithm to sequentially identify the top $\bar{L} = 100$ delay-angle pairs $\left\{\tau_l, \varphi_l\right\}_{l=1}^{\bar{L}}$ that best match the observation, based on a delay–angular domain dictionary constructed via uniform sampling over $\mbf{a}_N(\cdot) \otimes \mbf{b}_M(\cdot)$.
	Then, the coefficients $\{\beta_l \sqrt{\rho_l}\}_{l=1}^{\bar{L}}$ are estimated by LS.
	
	\item \textbf{Element-wise MMSE \cite{ma2015turbo_cs}:} This method adopts an i.i.d. Bernoulli-Gaussian prior for the channel coefficients in the same transform domain used by OMP, based on which an element-wise MMSE denoiser is implemented.
	The prior sparsity level is set to $0.15$.
	\item \textbf{Variational line spectral estimation (VALSE) \cite{valse2017tsp}:} A super-resolution algorithm that computes the approximate posterior pdfs of the delays, angles, and path coefficients.
	\item \textbf{Genie-aided lower bound:} This baseline assumes perfect knowledge of the path powers, delays, and angles $\left\{\rho_l, \tau_l, \varphi_l\right\}_{l=1}^{L}$,
	and employs the MMSE estimator to recover the complex coefficients $\left\{\beta_l\right\}_{l=1}^L$.
\end{itemize}

\begin{figure}
	[t]
	\centering
%		\vspace{-1em}
	\includegraphics[width=.95\columnwidth]{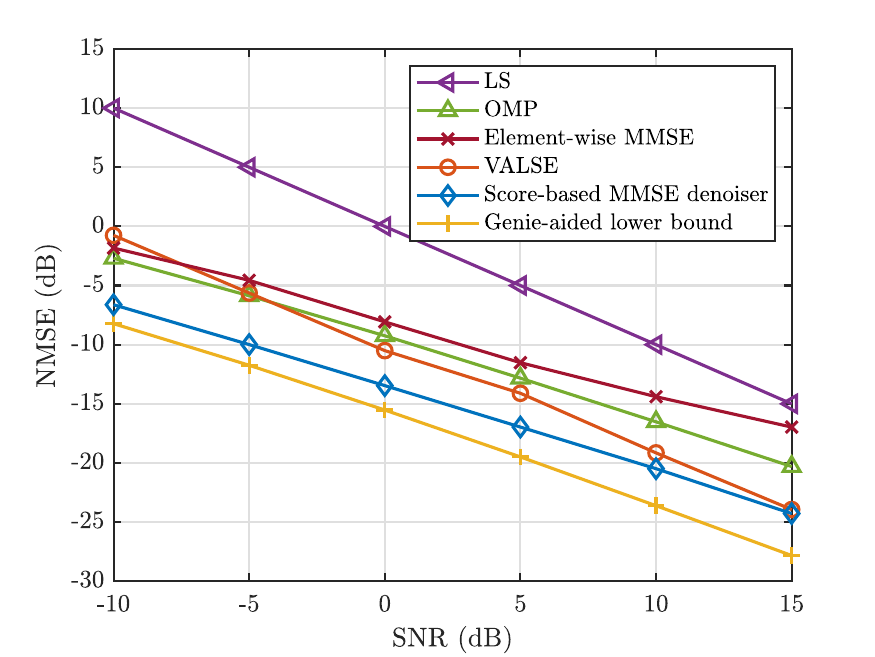}
%		\vspace{-.5em}
	\caption{Comparison of the denoising performance under AWGN by different denoisers.}
	\label{fig5}
%		\vspace{-.5em}
\end{figure}

Fig. \ref{fig5} compares the denoising performance of the score-based MMSE denoiser against these baselines.
Remarkably, the score-based MMSE denoiser outperforms VALSE \cite{valse2017tsp}, the state-of-the-art line spectral estimation algorithm, and closely approaches the unattainable genie-aided low bound.
Our score-based MMSE denoiser requires only a forward propagation through the first-order score network.
In contrast, the VALSE algorithm suffers from prohibitively high computational complexity due to the iterative evaluation of continuous pdfs, making it less suitable for real-time channel estimation.

\begin{figure}
	[t]
	\centering
%		\vspace{-1em}
	\includegraphics[width=.95\columnwidth]{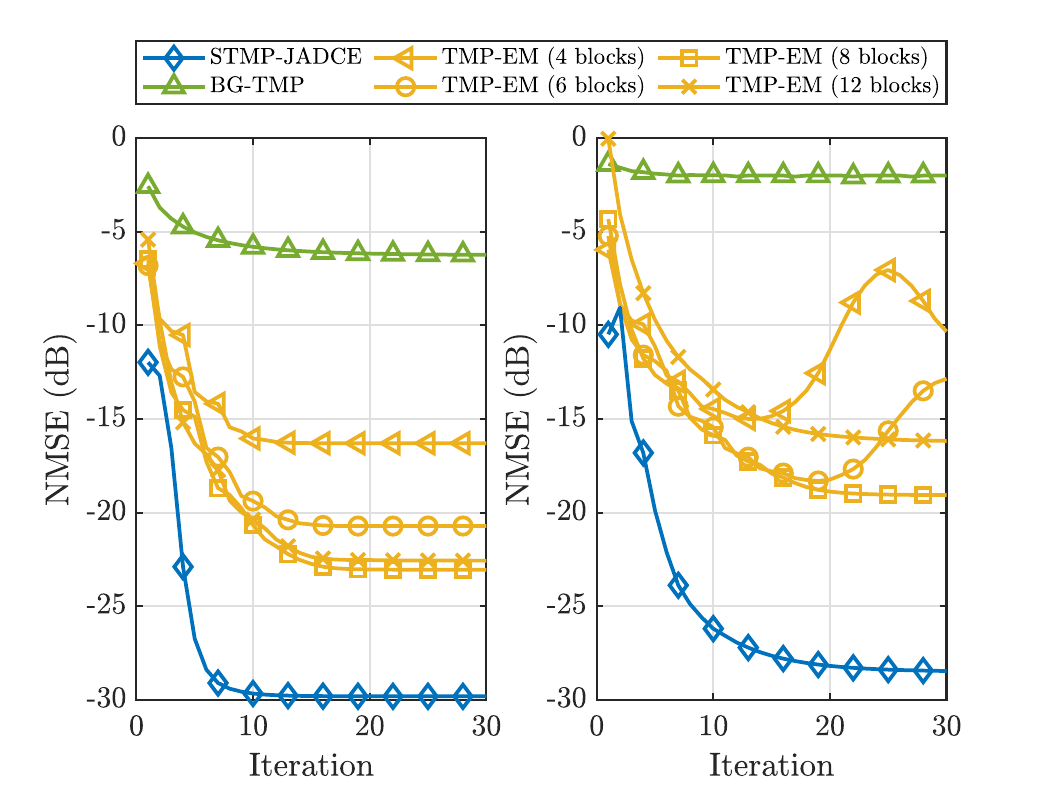}
%		\vspace{-.5em}
	\caption{NMSE in channel estimation versus the iteration number, where $\lambda = 0.05$, $T=30$, and the SNR is $10$ dB.
		Left: $K=800$ and we set the damping factor $\gamma = 0.8$ for all schemes;
		right: $K=1600$ and we set $\gamma = 0.6$ for all schemes. }
	\label{fig_convergence}
%		\vspace{-.5em}
\end{figure}

\subsection{JADCE Performance}
We compare the proposed STMP-JADCE algorithm against the following baselines:
\begin{itemize}
	\item \textbf{Bernoulli-Gaussian TMP (BG-TMP):} This method adopts the same TMP framework in Algorithm \ref{alg:framework}, but incorporates an i.i.d. Gaussian prior for channel denoising.
	\item \textbf{Generalized MMV-AMP (GMMV-AMP) \cite{gaozhen2020tsp}:}
	This algorithm exploits the structured sparsity of the channel matrix observed at multiple receive antennas and multiple subcarriers.
	The structured sparsity arises from the sporadic access of devices.
	\item \textbf{TMP-EM \cite{jiang2022turbo_mp}:}
	This algorithm performs JADCE based on a block-wise linear model in the frequency domain.
	The EM algorithm \cite{BiGAMP2014tsp} is applied to learn the prior variance of the parameters of the block-wise linear model.
	In TMP-EM, the number of sub-blocks is a hyperparameter that balances model accuracy with the number of free variables.
\end{itemize}

Fig. \ref{fig_convergence} shows the convergence behaviors of STMP-JADCE compared to various baselines.
For all plots in the left subfigure ($K=800$), we use a damping factor of $\gamma = 0.8$, while in the right subfigure ($K=1600$), we apply a damping factor of $\gamma = 0.6$.
Under this setting, GMMV-AMP frequently fails to converge and is therefore omitted from the results.
When $K=800$, STMP-JADCE converges within $10$ iterations, outperforming TMP-EM and BG-TMP both in convergence speed and NMSE performance.
As $K$ increases to $1600$, all the schemes experience a slight decrease in convergence speed.
Nevertheless, our method still converges in approximately $20$ iterations.
In contrast, TMP-EM shows substantial fluctuations when configured with either $4$ or $6$ linear sub-blocks.
We also note that the second iteration in STMP-JADCE leads to an increase in NMSE.
We conjecture that this is due to substantial estimation errors in the large-scale fading component during the initial iterations.
This causes a power mismatch in the input to the score networks, ultimately degrading the denoising performance.

\begin{figure}
	[t]
	\centering
	%		\vspace{-1em}
	\includegraphics[width=.95\columnwidth]{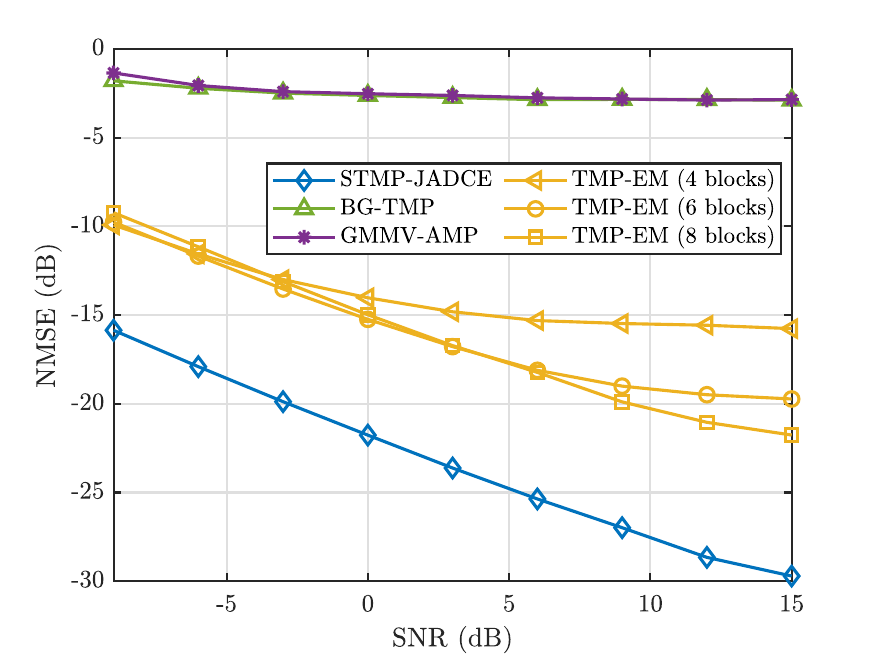}
	%		\vspace{-.5em}
	\caption{NMSE in channel estimation against SNR, where $\lambda = 0.05$, $K= 800$, and $T=20$.}
	\label{fig1_nmse}
	%		\vspace{-.5em}
\end{figure}

\begin{figure}
	[t]
	\centering
	%		\vspace{-.1em}
	\includegraphics[width=.95\columnwidth]{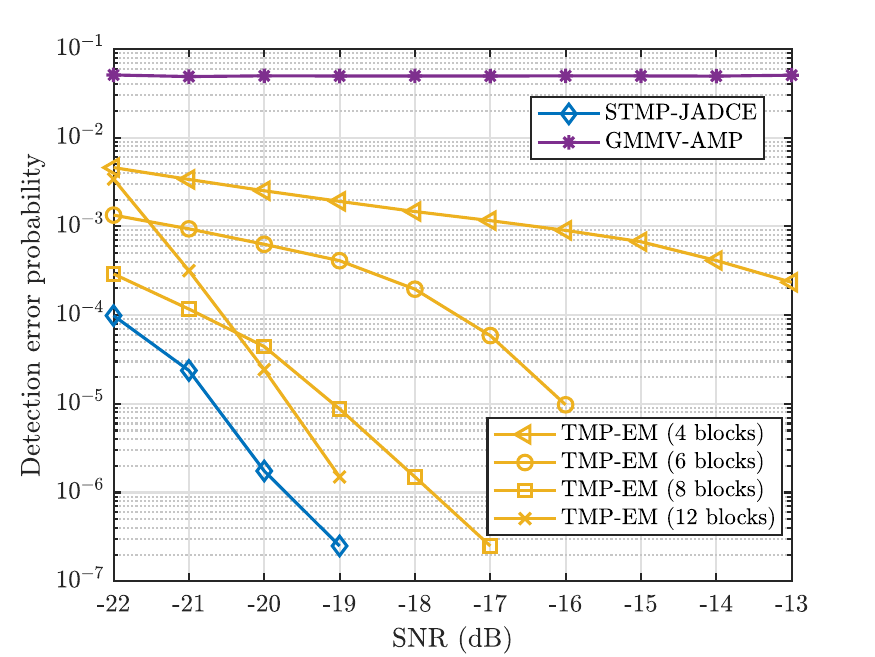}
	%		\vspace{-.5em}
	\caption{Detection error probability against SNR, where $\lambda = 0.05$, $K= 800$, and $T=20$.}
	\label{fig1_pe}
	%		\vspace{-.5em}
\end{figure}

Fig. \ref{fig1_nmse} shows the NMSE in channel estimation versus the SNR.
Our method demonstrates a significant improvement over the baselines across all SNRs.
The performance gap between our method and BG-TMP highlights the critical importance of an accurate channel prior in message passing.
Moreover, the TMP-EM algorithm favors a smaller number of sub-blocks at low SNR, while preferring a larger number of sub-blocks at high SNR.
The sensitivity to the number of sub-blocks can pose challenges in hyperparameter tuning.
In contrast, STMP-JADCE learns the channel prior directly from data, eliminating the need for hand-crafted modeling and achieving superior adaptability with consistent performance gains over model-based baselines.

\begin{figure}
	[t]
	\centering
	%		\vspace{-1em}
	\includegraphics[width=.95\columnwidth]{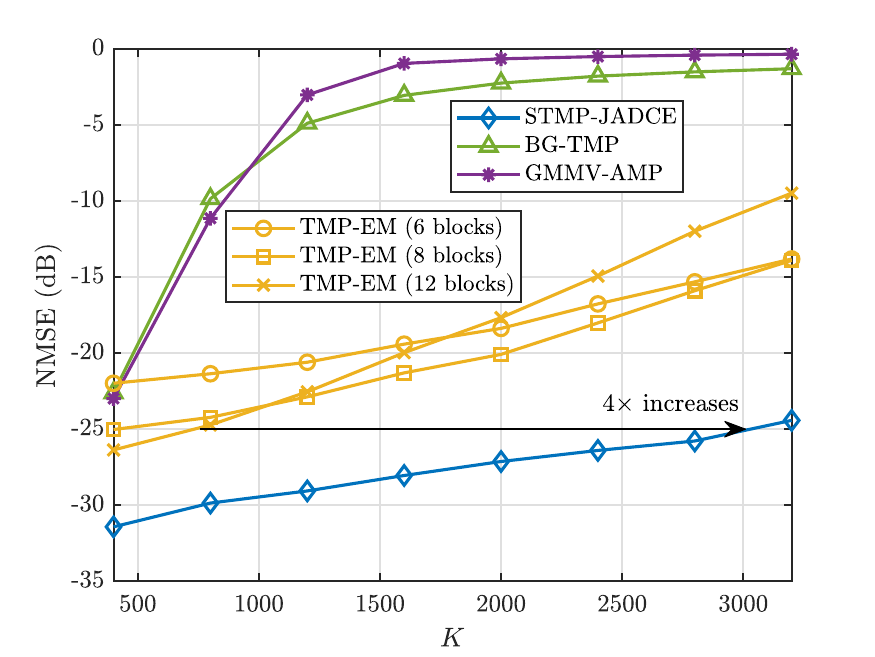}
	%		\vspace{-.5em}
	\caption{NMSE in channel estimation against the number of devices $K$, where $\lambda = 0.05$, $T=40$, and the SNR is $10$ dB.}
	\label{fig2_K}
	%		\vspace{-.5em}
\end{figure}

In Fig. \ref{fig1_pe}, we present the detection error probability, which sums up missed detections and false alarms, plotted against different SNRs.
The BG-TMP algorithm often fails to converge and is not included.
Interestingly, although all schemes utilize the same Bernoulli prior for device activity, our method achieves the lowest detection error probability.
This improvement can also be attributed to the plug-in of the score-based generative channel prior.
The reason is the following.
As shown in Fig. \ref{fig1_nmse}, the accurate channel prior brings a remarkable gain in channel estimation.
Through message-passing iterations, the device activity is detected using this more accurate channel estimate, hence resulting in enhanced detection performance.

\begin{figure}
	[t]
	\centering
%	\vspace{-.5em}
	\includegraphics[width=.95\columnwidth]{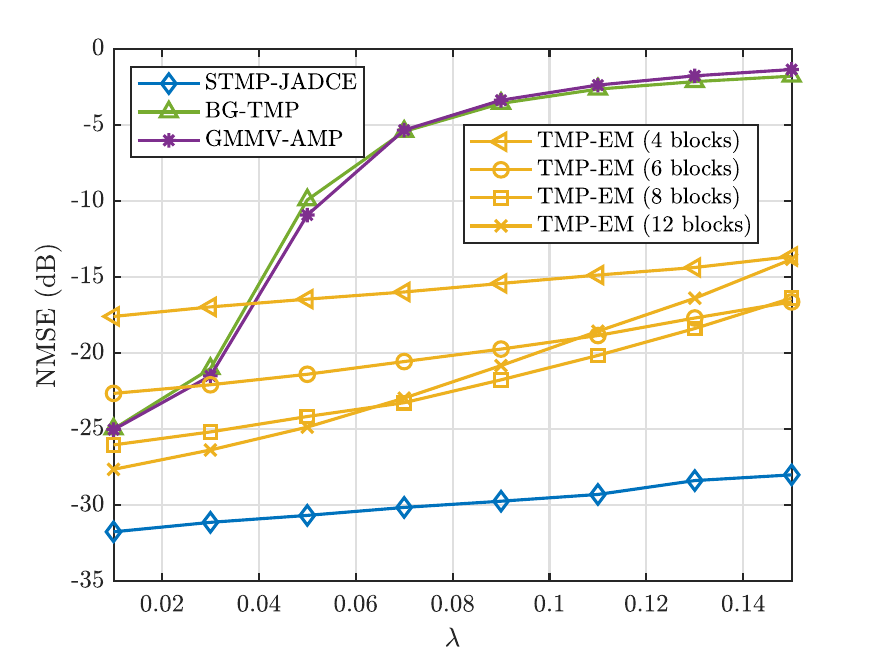}
%	\vspace{-.5em}
	\caption{NMSE in channel estimation against the device active probability $\lambda$, where $K= 800$, $T=40$, and the SNR is $10$ dB.}
	\label{fig3_lambda}
%	\vspace{-.5em}
\end{figure}

In Fig. \ref{fig2_K}, we plot the achieved NMSE in channel estimation for various numbers of total devices. 
Targeted at a channel estimation NMSE of $-25$ dB, our method can support $3000$ connected devices, while the best baseline (TMP-EM with $12$ sub-blocks) can accommodate only $750$ devices, leading to a $4\times$ increase in access capacity.
Moreover, for NMSE values below $-5$ dB, STMP-JADCE shows the flattest slope in the performance curve. 
This suggests that, given an equivalent tolerance in NMSE degradation (in dB), our approach can support the largest increase in the number of connected devices.
These observations highlight the superiority of the proposed algorithm for achieving massive connectivity.

\begin{figure}
	[t]
	\centering
%		\vspace{-1em}
	\includegraphics[width=.95\columnwidth]{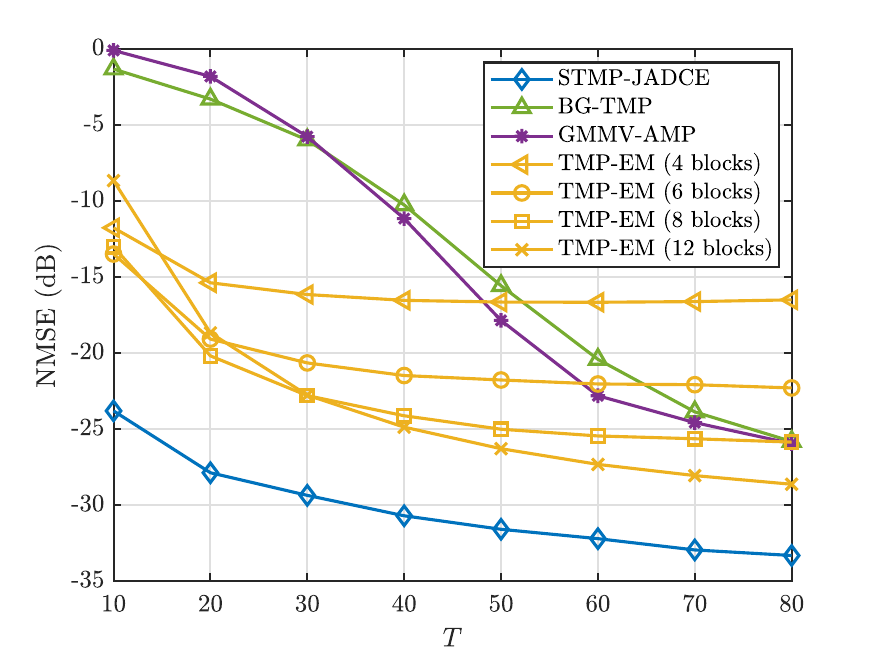}
%		\vspace{-.5em}
	\caption{NMSE in channel estimation against the pilot sequence length $T$, where $\lambda = 0.05$, $K= 800$, and the SNR is $10$ dB.}
	\label{fig4_T}
%		\vspace{-.5em}
\end{figure}

Fig. \ref{fig3_lambda} shows the NMSE performance by varying the device active probability from $0.01$ to $0.15$.
It is observed that STMP-JADCE consistently outperforms all other schemes by a large margin.
In addition, the NMSE of STMP-JADCE deteriorates more gracefully than that of the other schemes as the active probability increases.
Fig. \ref{fig4_T} investigates the influence of pilot sequence length on the NMSE in channel estimation.
The proposed algorithm has a roughly $5$ dB gain in NMSE when $T=80$, and the gain is increased to $10$ dB when $T=10$.
This demonstrates the distinct advantage of the proposed method in the low pilot overhead regime.

\begin{figure}
	[t]
	\centering
	%		\vspace{-1em}
	\includegraphics[width=.95\columnwidth]{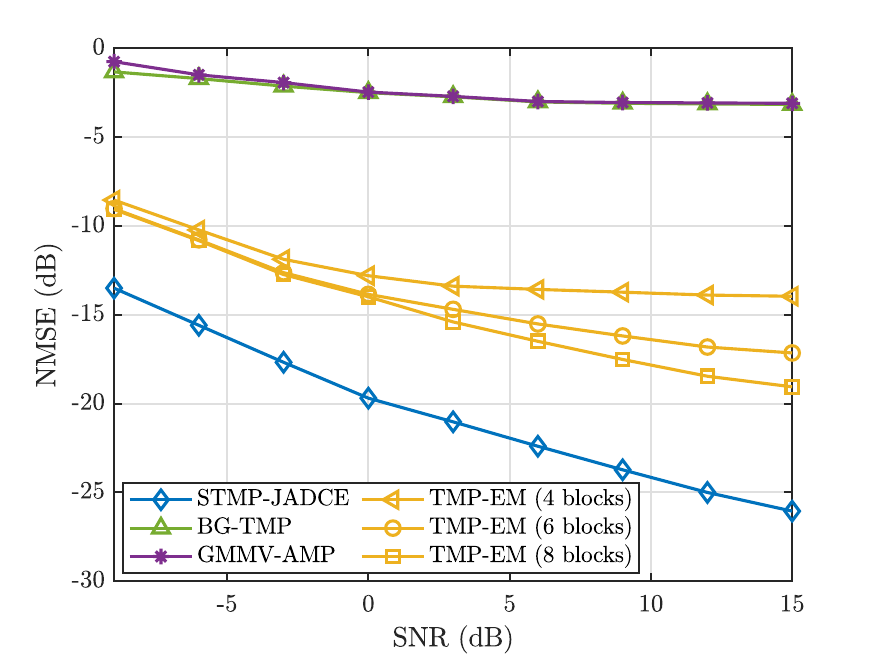}
	%		\vspace{-.5em}
	\caption{NMSE in channel estimation versus SNR for cross-CDL evaluation, where $\lambda = 0.05$, $K= 800$, and $T=20$.
		The score networks are trained on the CDL-C dataset, while the STMP-JADCE algorithm is evaluated on the distinct CDL-B channel model.
		}
	\label{fig_major_cdl_b}
	%		\vspace{-.5em}
\end{figure}

\begin{figure}
	[t]
	\centering
	%		\vspace{-1em}
	\includegraphics[width=.95\columnwidth]{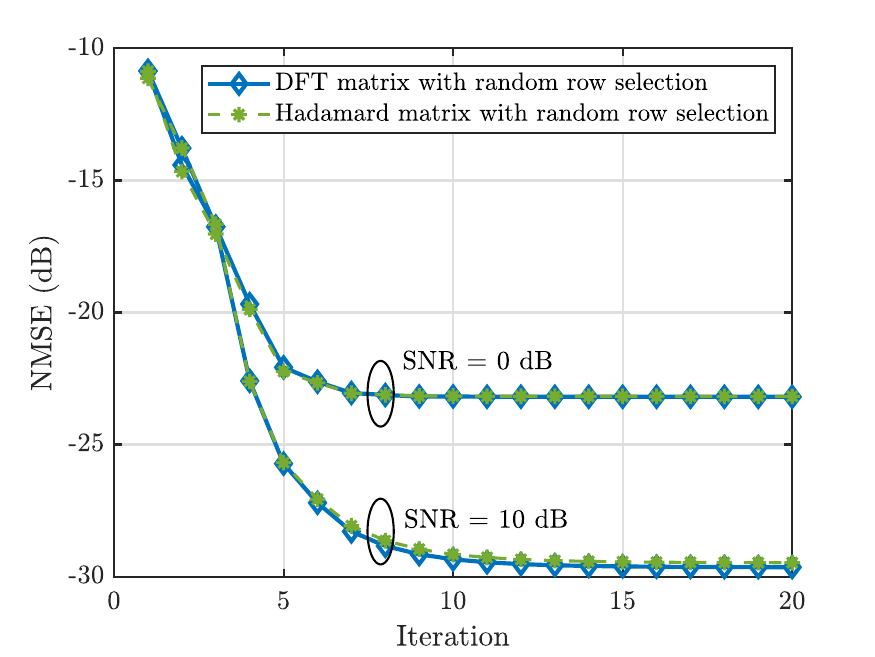}
	%		\vspace{-.5em}
	\caption{Convergence behaviors of STMP-JADCE under different measurement matrices, where $\lambda = 0.05$, $K= 1024$, $T = 30$, and $\gamma = 0.8$.}
	\label{fig_major_hadamard}
	%		\vspace{-.5em}
\end{figure}

\begin{figure}
	[t]
	\centering
	%		\vspace{-1em}
	\includegraphics[width=.95\columnwidth]{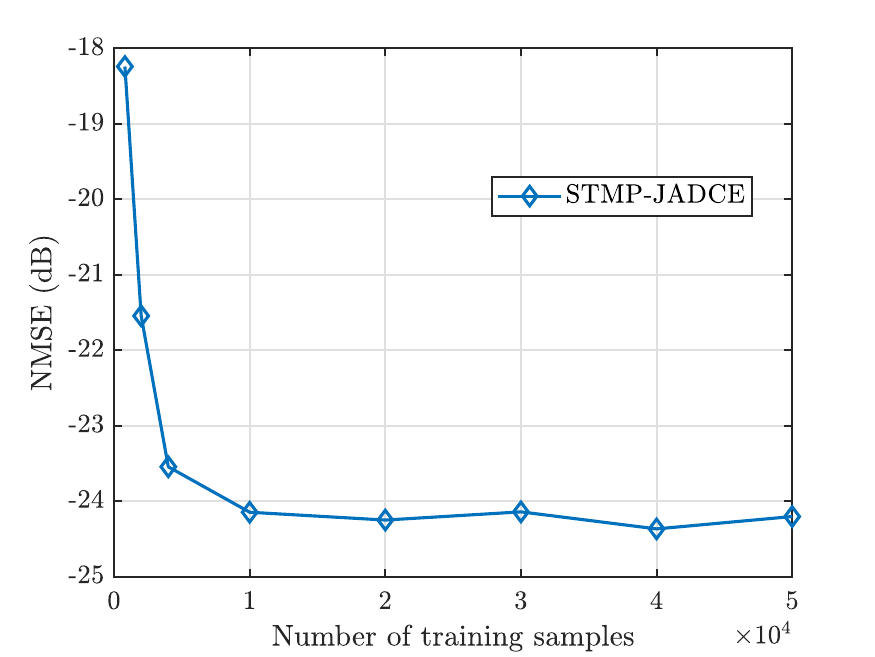}
	%		\vspace{-.5em}
	\caption{NMSE in channel estimation versus the number of training samples, where $\lambda = 0.05$, $K= 800$, $T = 30$ and the SNR is $3$ dB.}
	\label{fig_major_scarcity}
	%		\vspace{-.5em}
\end{figure}

To examine the robustness of STMP-JADCE under channel model mismatch, we conduct a cross-CDL evaluation in which the score networks are trained on the CDL-C dataset, while the STMP-JADCE algorithm is evaluated on the distinct CDL-B channel model.
As shown in Fig.~\ref{fig_major_cdl_b}, STMP-JADCE maintains strong performance and continues to outperform all baselines across the entire SNR range. 
This cross-model consistency can be attributed to the randomized generation of angles, delays, and complex path gains in the CDL-C training set, 
which exposes the score prior to a broad variety of channel realizations, 
as well as to the inherent generalization capability of the generative score networks.

We further evaluate STMP-JADCE under different measurement matrices without retraining the generative prior.
In addition to the DFT-based design, we consider a Hadamard pilot matrix whose entries are binary $\{\pm1\}$ and mutually orthogonal across rows, providing a real-valued and multiplication-free alternative to complex-valued DFT pilots.
As shown in Fig.~\ref{fig_major_hadamard}, the algorithm exhibits consistent convergence behavior and comparable NMSE performance across DFT and Hadamard pilots.
This consistency arises naturally from the plug-and-play formulation of STMP-JADCE, in which the unconditional score prior operates independently of the pilot structure, allowing seamless adaptation to various pilot designs drawn from suitably random matrix ensembles.

Finally, to assess the data efficiency of the proposed STMP-JADCE, we evaluate its performance under different training set sizes in Fig.~\ref{fig_major_scarcity}.
The performance rapidly improves as the dataset size increases from $8 \times 10^2$ to $4\times 10^3$, and then quickly saturates beyond $1\times10^4$ samples.
This result indicates that STMP-JADCE requires only a moderate amount of data to achieve near-optimal performance, demonstrating strong data efficiency and robustness even when large-scale datasets are unavailable.

\section{Conclusion and Discussion} \label{sec:conclusions}
In this paper, we introduced STMP-JADCE for massive connectivity, a model-based deep learning solution that combines message passing with score-based generative models.
In each message-passing iteration, our approach maintains both a first-order and a second-order score network to calculate the posterior mean and variance of the channel, respectively.
We show by simulation results that STMP-JADCE drastically outperforms all competing algorithms, while achieving a comparable or even faster convergence speed.

From a more general perspective, this work evidences that with carefully specified message update rules and approximations,
message passing allows for the integration of unconditional score-based generative models to solve Bayesian inference problems without requiring task-specific training.
This opens up new opportunities for the STMP framework to tackle a variety of inverse problems arising in communication, sensing, and other application domains.

\bibliographystyle{IEEEtran}
\bibliography{IEEEabrv,mybib}

\end{document}